\documentclass[final,twocolumn,twoside,journal]{IEEEtran}
\usepackage{amsmath,amssymb}
\usepackage[level=3]{wgroup_message}  
\usepackage{hyperref} 
\usepackage{acronym}  
\usepackage{color}  
\usepackage{graphicx,psfrag,cite,subfigure}
\usepackage{enumerate}

\usepackage[table]{xcolor}
\definecolor{lightgray}{gray}{0.9}

\usepackage{amsthm} 
\usepackage{amsfonts}
\usepackage{amsmath}
\usepackage{float}

\theoremstyle{definition}
\newtheorem{theorem}{Theorem}

\newtheorem{corollary}{Corollary}

\newtheorem{example}{Example}

 \acrodef{SE}{state estimation} 
 \acrodef{TL}{transmission line} 
 \acrodef{PMU}{phase measurement units}
 \acrodef{SCADA}{supervisory control and data acquisition}
 \acrodef{MSE}{mean squared error}
 \acrodef{MMSE}{minimum-mean square estimate}
 \acrodef{HMM}{Hidden Markov Model}
 \acrodef{UKF}{uncented Kalman filter}
 \acrodef{EKF}{extended Kalman filter}
 \acrodef{BCF}{belief condensation filter}
 \acrodef{PF}{particle filtering}
 \acrodef{LSQ}{least squares estimation}

\DeclareMathOperator*{\argmax}{arg\,max}

\newcommand{\bd}{\begin{description}}
\newcommand{\ed}{\end{description}}
\newcommand{\be}{\begin{enumerate}}
\newcommand{\ee}{\end{enumerate}}
\newcommand{\bi}{\begin{itemize}}
\newcommand{\ei}{\end{itemize}}
\newcommand{\bl}{\begin{list}}
\newcommand{\el}{\end{list}}
\newcommand{\bt}{\begin{tabbing}}
\newcommand{\et}{\end{tabbing}}

\newcommand{\paperTitle}{State Estimation for Future Energy Grids}

\begin{document}

{



\twocolumn

}


\title{\paperTitle}


\author{
	\vspace{0.2cm}
    Shervin~Mehryar,~\IEEEmembership{Member,~IEEE},
    Moe~Z.~Win,~\IEEEmembership{Fellow,~IEEE}
    \\\vspace{0.3cm} 
    Laboratory for Information and Decision Systems (LIDS)\\
    Massachusetts Institute of Technology (MIT)\\
    77 Massachusetts Avenue, Room 32-D674A\\
   Cambridge, MA 02139 USA\\
    Tel.: (617) 253-9341\\
    e-mail: {\tt moewin@mit.edu}
        }

\maketitle 

\markboth{}{}


\setcounter{page}{1}

\begin{abstract}
Today's power generation and distribution networks are quickly moving toward automated control and integration of renewable resources - a complex, integrated system termed the Smart Grid. A key component in planning and managing of Smart Grids is State Estimation (SE). The state-of-the art SE technologies today operate on the basis of slow varying dynamics of the current network and make simplifying linearity assumptions. However, the integration of smart readers and green resources will result in significant non-linearity and unpredictability in the network. Therefore in future Smart Grids, there is need for ever more accurate and real-time algorithms. In this work, we propose and examine new SE methods that aim to achieve these measures by approximating the true distribution of the state variables, rather than a linearized version as done for instance in Kalman filtering. Through simulations we show that in the presence of non-linearities and non-Gaussian noise, our general SE framework improves accuracy where linear and Kalman-like filters exhibit impaired performance.
\end{abstract}


\section{Introduction}

Operation, monitoring, and control of power grids rely on the \ac{SE} problem, defined as the computation of line voltages and phase angles under steady-state conditions \cite{AbuExp:04}. Estimated line parameters are utilized by supervisory entities to ensure safe-mode operation, efficient resource allocation, and low-cost power generation, among others. While traditionally energy networks are assumed to have slow and modest behavior, in future girds real-time dynamic \ac{SE} is essential. Above all, for a expanding system that will heavily rely on estimations at every level, accuracy becomes paramount with minimal tolerance for error. This work focuses on state estimation methods that provide reliable and efficient performance with high scalability and low complexity cost for future grids.

\begin{figure}[t]
  \centering
  \includegraphics[width=0.48\textwidth,keepaspectratio]{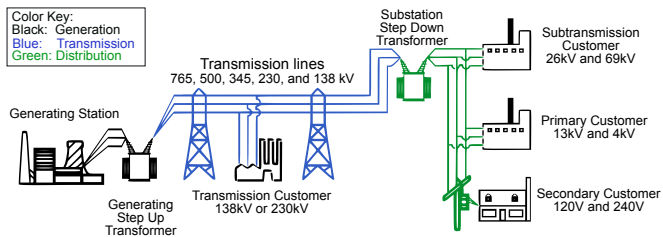}
  \caption{Electric power generation, transmission, and distribution diagram.}
  \label{fig:3layers}
\end{figure}

The power grid of today has a 3-layer structure: 1) power is generated at Generation Plants, 2) power is transferred through \ac{TL}, 3) power is delivered to end users at Distribution Level. Figure \ref{fig:3layers} \footnote{Image courtesy of: www.smartgridlegalnews.com/smart-grid-basics/the-electric-grid-101.} shows a diagram of the power flow through such a network. The Smart Grid in this context refers to all aspects of the power network that allow for smooth and seamless generation and distribution of power. It is, according to \cite{HuaWerHuaKasGup:12,Wu:90}, an automated, widely distributed energy delivery network, that will be characterized by a two-way flow of electricity and information and will be capable of monitoring everything from power plants to customer preferences and individual appliances. It incorporates into the grid the benefits of distributed computing and communications to deliver real-time information and enable the near-instantaneous balance of supply and demand at the device level. As opposed to a top-down flow, the Smart Grid will be highly integrated and dynamic. 

Traditionally, \ac{SE} only exists in static form \cite{HuaWerHuaKasGup:12,BroSou:09,GloSarOve:11,VanRib:83,GomVilGomRouVan:11}. This is mainly due to the fact that current power generation is facilitated primarily by hydro (e.g. water powered) and steam (e.g. coal or gas fueled) turbines that follow slow-varying and predictable patterns. However, with the advent of renewable sources such as wind and solar panels that are more irregular, the exigence in dynamic estimation becomes apparent.

In addition, traditional \ac{SE} focuses on \ac{TL} level estimation, with \ac{SE} at Distribution level happening very infrequently \cite{HuaWerHuaKasGup:12}. At the time, power meters that provide power and voltage magnitude readings on high-voltage lines, are the sole data acquisition source. A limiting factor here is the difficulty in deploying measurement units given the size of the network. Today however, affordable smart meters and \ac{PMU} are available that allow for more distributed (at households) and content-rich data collection. Particularly with the emergence of green sources as key contributors to future grids, power generation no longer will follow a vertical flow from plant generation down to house-hold consumption. This shift in generation trend puts distributed level \ac{SE} on par with \ac{TL} level \ac{SE}.

Table \ref{table:summary} summaries the current state of the grid versus the changes that are going to be seen in future energy networks. In view of these transformations, \ac{SE} algorithms must be able to deal with highly dynamic and heterogenous networks and operate reliably at Distributed Levels as well.

\begin{table}
\caption{Comparison of current versus future power grids.}
\label{table:summary}
\begin{center}
\rowcolors{2}{}{lightgray}
\begin{tabular}{| c | c |}\hline
\textbf{Traditional Girds} & \textbf{Future Grids} \\\hline 
Vertical Structure   & Integrated Structure \\
Static Estimation   & Dynamic Estimation \\
Transmission Level \ac{SE}   & Distributed \ac{SE} \\\hline
\end{tabular}
\end{center}
\end{table}

Optimal filtering methods, as they are often called, exist for finite state, linear, and Gaussian problems, as well as a constrained sub-class of non-linear problems. However, in general, continuous problems with non-linear/non-Gaussian models are intractable and approximate solutions that are only locally optimal are in application \cite{MazShiWin:13}.

\subsection{4 Key Design Considerations for Future Energy Networks}
Based on the discussion building up to here, there are four major considerations that are key to ensuring flawless and reliable design of the future smart grids. Any \ac{SE} algorithm must have taken these 4 issues into account and addressed them in order to meet future operational and practical requirements of the network. They are enumerated in what follows.

\subsubsection{Large Network Scalability} 

One of the key considerations in designing algorithms for the Energy Networks of tomorrow is the scale of such networks. Considering the U.S.A. power grid alone, It consists of more than 9,200 electric generating units with more than 1,000,000 megawatts of generating capacity connected through more than 300,000 miles of \ac{TL} \cite{SmartGrid:08,JakKez:02,Kor:11,ZhaAbu:05}. The network itself is subdivided into three main interconnections: Eastern, Western, and Texas. Each of which is built from countless number of electricity transmission components. For what is orders of magnitude larger than the Internet backbone in the number of building blocks, while at the same time lacking an inherent monitoring layer like the Internet network, the architecture and build of operation services becomes crucial. Given such a scale, highly distributed and precise estimation algorithms are of outmost necessity where taking measurements form every node in the network is unrealistic and a minor error can cascade through the Energy Network with catastrophic consequences.

\subsubsection{Integrated and Dynamic Nature}

One of the key goals in building the future Smart Grids according to the report by the U.S. Department of Energy is for the grid to be fully accommodating to both traditional and renewable sources of all sorts, where the power flow is no longer uni-directional from Generating Plants down to the Distribution Level. The Energy Network of tomorrow will be highly integrated and power injection can happen realistically at any point in the network, with smart \ac{PMU}'s collecting and sending feedback to the \ac{SCADA} centers at rates about 30 samples/second. As a result the future network will be highly dynamic and their state can change in unpredictable manners. This makes the need for adaptive and fast-response estimation algorithms indisputable.  

\subsubsection{Non-linear System Immunity}

A technical challenge in Energy Networks that is often overshadowed, is the presence of non-linearities at the low-level system. The network is built of active and reactive components that 
have complex form and beset network parameters with non-linear relations. For instance, the relation between voltage magnitudes and angles and the corresponding real and imaginary powers for a line are given by an equation of the form \cite{MacBiaBum:11}:
\begin{align}
 P_{ij} &= |V_i|^{2} (g_{i0} + g_{ij}) \\ &-  |V_i| |V_j| (  g_{ij} \cos( \theta_i - \theta_j ) - b_{ij} \sin( \theta_i - \theta_j ) ) \nonumber \\
 Q_{ij} &= - |V_i|^{2} (b_{i0} + b_{ij}) \\ & -  |V_i| |V_j| (  g_{ij} \sin( \theta_i - \theta_j ) - b_{ij} \cos( \theta_i - \theta_j ) )  \nonumber 
\end{align}
Such non-linearities have irreprehensible precautions for estimation algorithms and can degrade performance if not properly dealt with. Most state-of-the-art methods unfortunately are designed only to work well disregarding such conditions.

\subsubsection{Uncategorized System Noise}

Lastly, a major hurdle to computing the exact line states is the uncertainty in the network. There is noise arising at all levels of the system, from measurement units and meters to data fusion at the control centers. Given the scale of the network, this predicament can escalade very rapidly. To add salt to the injury, the exact form of the system noise is not known. Therefore, estimation algorithms must have built-in resilience to uncategorized noise models or at the very least mitigate the effect of uncertainty through other intelligent ways.

\subsection{Review of State of the Art} \label{sec:soa}

In view of the network considerations discussed above, many recent studies and developments have been conducted to realize the possibility of green, adaptive, integrated Smart Girds. At the state estimation level, there are a few key components in the system that have been subject to extensive research. In this section, a review of systematic shortcomings related to these components is presented and the progress in addressing them is discussed. 

The problem of \ac{SE} can be cast as a \ac{HMM} whereby first the state of the system is predicted based on previous states (Prediction Step) and next the predictions are corrected by taking into account the network measurements (Update Step). Mathematically, the problem can be described by the computation of a state vector $\mathbf{x}_t$ at time $t$ from the previous state through a function $g(\cdot)$ and given a set of measurements $\mathbf{y}_t$ through a function $h(\cdot)$.  For instance, the state vector can be line voltages and measurements can be a set of power, voltage, and current readings taken by a batch of \ac{PMU}'s. The equations that govern these relations are:
\begin{align}
\mathbf{x}_{t+1}= g( \mathbf{x}_t) + \mathbf{q}_t  \\
\mathbf{x}_t= h(\mathbf{y}_t) + \mathbf{n}_t 
\end{align}
known as the Dynamic Model and Observation Model in the literature. Here, $ \mathbf{q}_t $ and $ \mathbf{n}_t $ are related to the uncertainty in each model.

Three points of active research in this field are:

\begin{enumerate}[i)]

	\item There exists a lack of a physically and mathematically sound dynamic model. Current models assume smooth and slow state progression \cite{GhaKam:11, PenHuaSunDiaKalAndLiLee:11, ValTer:11}. For instance, Holt's Smoothing method is a widely used linear Dynamic Model given by: 
	\begin{align}
		\mathbf{x}_{t+1}= \mathbf{F}_t \mathbf{x}_t + \mathbf{g}_t + \mathbf{q}_t \label{eq:assumption1}
	\end{align}
	In the future, the addition of renewable energy inlets throughout the system and automated network controllers will make the system significantly more unpredictable and fluctuating. There is therefore a need for a model that meets the requirements of future Smart Grids in that it can support the Integrated and Dynamic nature of the system.\\
	
	\item The measurement equation $h(\cdot)$ in Power Systems are highly non-linear and complex. Current methods, such as Kalman-like filters \cite{GhaKam:11, PenHuaSunDiaKalAndLiLee:11, ValTer:11}, assume linearization which impairs the accuracy. For instance, line voltage and power relation discussed previously are often simplified as:
	\begin{align}
		P_n = \sum_{i \neq n} b_{ni} (\theta_n - \theta_i) \label{eq:assumption2}
	\end{align}
	There is a need for filtering methods that can promise non-linear system immunity. \\

	\item With increasing deployment of wireless smart readers and \ac{PMU}'s, more complicated uncertainty modeling is needed. Typically measurement noise is assumed to be additive Gaussian in all current literature, in the form:
	\begin{align}
		\mathbf{n}_t \sim \mathcal{N}(\mathbf{0}, \mathbf{\Sigma}_{n}^t) \label{eq:assumption3}
	\end{align}
	Since the true nature of the noise is not known yet, there is a need for better models that can capture more complicated, uncharted system uncertainty.
	
\end{enumerate}

The goal here is to develop estimation algorithms customized to operate in accordance with the requirements of future power networks. In view of the discussions above, this work is after algorithms that are distributed, resilient to different noise formations, and adapt to different system dynamics. This work proposes methods that particularly focus on reducing estimation error in the face of non-linearities and non-gaussian noise. Much recent research has gone into addressing these aspects and a summary of most popular and up-to-date methods is provided in Table \ref{table:stateofart}. All these related studies make simplifying assumptions for the Observation Model that rely on linearization and Gaussian representation.

\begin{table*}
\begin{center}
\caption{Summary of state-of-the-art filtering methods.} 
\begin{tabular}{ || p{5cm} | p{3cm} | p{5cm} ||}\hline
Filtering Technique & Families of Distributions & Filtering Approach \\\hline \hline
Least Squares Minimization (LSQ) \cite{SilFilQue:83}  \cite{DebLar:70} \cite{BarKel:94} & N/A & Minimize objective function by Newton�s Method  \\\hline
Kalman Filtering (KF), Extended Kalman Filtering (EKF) \cite{GhaKam:11} \cite{PenHuaSunDiaKalAndLiLee:11} \cite{ChoChoParChu:09} \cite{BiQinYan:08} \cite{MonGar:83}  \cite{WanSch:04} & Gaussian Distribution  & Linearize the models by Taylor Expansion \\\hline
Unscented Kalman Filtering (UKF) \cite{ValTer:11}  & Gaussian Distribution & Approximate KF recursions by numerical integration with different quadrature rules  \\\hline
Belief Condensation Filtering (BCF) \cite{mehryar2019belief} & Mixture of Exponential Families (e.g. Gaussian) & Approximate complex distributions by right sided Kullback-Leibler (KL) divergence \\\hline
Particle Filtering (PF) \cite{AruMasGorCla:02} & Discrete Distributions & Approximate complex distributions by random sampling with different proposal densities \\\hline
\end{tabular} \label{table:stateofart}
\end{center}
\end{table*}

This work presents two well-known filtering methods, namely the \ac{BCF} and the \ac{PF} which have been used in other domains such as Network Localization and Navigation \cite{WymLieWin:09}, and applies them to the problem of \ac{SE} in Energy Networks. These are recent filtering techniques that perform provably well under non-linear, non-gaussian conditions. The effect of dynamic modeling is left as a subject of future research in this work. To this effect, a forecasting method that is commonly used in the literature is also employed here. In Section \ref{sec:PowerSystem}, a review of the Power Systems Theory and the complex equations relating observed and state variables are presented. In Section \ref{sec:PowerSystem}, the \ac{HMM} used for modeling and forecasting in power networks is explained. Detailed descriptions along with preliminary analysis of the proposed filtering methods (the \ac{BCF} and \ac{PF}) are provided in Sections \ref{subsec:BC} and \ref{subsec:PF}. In Section \ref{chap3}, in-depth numerical experimentation is carried out to analyze the performance of the \ac{BCF} and the \ac{PF}. The simulation set-up is described in Section \ref{sec:SimSetup}. The characterization of non-linear versus linear filtering in power networks is provided in Section \ref{sec:LandNLF}. Section \ref{sec:Noise} details out the presence and effect of non-Gaussian uncertainty in the network. A full-system simulation that measures and compares the performance of pertinent algorithms in a real-time scenario is given in Section \ref{sec:FullSim}, along with complexity analysis in Section \ref{Appx:cplx}. Final conclusions and remarks are drawn in Section \ref{chap4}.

\section{System Model and Methodology} \label{chap2}

In this section, a review of the relevant parts of the Power Systems theory is presented with elaboration for their utilization within a \ac{HMM} framework for \ac{SE}. Based on this set-up, the development of the \ac{BCF} and \ac{PF} for the task of filtering over \ac{HMM}'s is detailed out as well.

\subsection{Power Systems} \label{sec:PowerSystem}

The network theory and equations for power systems have long been established \cite{SchWil:70,Sch:70}. Figure \ref{fig:14bus} shows a simple diagram of a power microgrid. Any such system consists of 4 primary system components: 1) Bus lines - which transfer power; 2) Branches - as connections between bus lines; 3) Generators - as sources of power; 4) Loads - as sinks for power. As far as network control goes, the state of generators are fully known to the operators. We are only concerned with state estimation for bus, branch, and load elements at different points, known as nodes in the network. The parameters of interest at each node to compute are complex voltages, powers, and currents (also called phasors), as summarized in Table \ref{table:params} for an $n$-node network. 

For the purposes of \ac{SE} in energy networks, it suffices to compute line voltage magnitudes and phases only. Given physical values (e.g. line resistances, impedances, and admittances), currents and powers can be calculated thereby. Put in Estimation Theory terminology, the complex voltage values form sufficient statistics, the knowledge of which tells us everything needed to know about the network. As such, the interesting part is the parameter relations in terms of voltage magnitudes and angles. The following sections are aimed at explaining these relation, which form the basis for \ac{SE}'s Observation Model in this work's analysis \cite{AbuExp:04}:  

\begin{table}
\caption{Power system parameters for network of size $n$.}
\label{table:params}
\begin{center}
\rowcolors{2}{}{lightgray}
\begin{tabular}{| l |c|}\hline
 \multicolumn{2}{|c|}{\textbf{Parameters}} \\\hline 
Bus Voltage Magnitudes  & $|V_1|, |V_2|, \cdots, |V_n|$  \\
Bus Voltage Phase Angles   & $\theta_1, \theta_2, \cdots, \theta_n$ \\
Bus Real Powers  & $P_1, P_2, \cdots, P_n$ \\
Bus Reactive Powers  & $Q_1, Q_2, \cdots, Q_n$ \\
Bus Current Magnitudes  & $|I_1|, |I_2|, \cdots, |I_n|$  \\
Bus Current Phase Angles   & $\angle{I_1}, \angle{I_2}, \cdots, \angle{I_{n}} $ \\\hline
\end{tabular}
\end{center}
\end{table}

\subsubsection{AC Power Flow Model}
A single-phase AC model for injected power at bus $n$ is given by:
\begin{align}
P_n &= \sum_{i \neq n} |V_i||V_n| ( g_{in} \cos(\theta_i - \theta_n )  \nonumber \\ &+ b_{in} \sin(\theta_i - \theta_n ) )\\
Q_n &= \sum_{i \neq n} |V_i||V_n| ( g_{in}  \sin( \theta_i - \theta_n) \nonumber \\  &- b_{in} \cos( \theta_i - \theta_n ) ) 
\end{align}
for the real power and reactive  power. Here, the summation is over all other connected buses to bus $n$, $ g_{ij}$ denotes conductance between bus $i$ and $j$, and
$ b_{ij}$ denotes susceptance between bus $i$ and $j$. A subscript $0$ denotes the reference (e.g. ground). The real and reactive power flow from bus $i$ to bus $j$ takes the form:
\begin{align}
P_{ij}  & = |V_i|^{2} (g_{i0} + g_{ij}) \nonumber \\  & -  |V_i| |V_j| (  g_{ij} \cos( \theta_i - \theta_j )  - b_{ij} \sin( \theta_i - \theta_j ) )  \label{eq:h_P}  \\
 Q_{ij} & = - |V_i|^{2} (b_{i0} + b_{ij}) \nonumber \\ & -  |V_i| |V_j| (  g_{ij} \sin( \theta_i - \theta_j )  - b_{ij} \cos( \theta_i - \theta_j ) ) \label{eq:h_Q}    
\end{align}

\subsubsection{Line Current and Voltage Model}
The relation between voltage-phasor measurements and voltage estimations can be expressed by a linear model. That is, a measured voltage value is a noisy version of the actual voltage state value. Unlike voltage measurements, voltage estimations form a non-linear relation with current-phasors. This relation using a $\pi$-model is given by:  
\begin{align}
\mathfrak{Re}\{I_{ij}\} & = (g_{ij}+g_{i0}) |V_i| \cos( \theta_i ) \nonumber \\ & - g_{ij} |V_j| \cos( \theta_j ) - (b_{ij}+b_{i0}) |V_i| \sin(\theta_i) \nonumber \\ & - b_{ij} |V_j| \sin( \theta_j )  \label{eq:h_I1} \\ 
\mathfrak{Im}\{I_{ij}\} & = (g_{ij}+g_{i0}) |V_i| \sin( \theta_i ) \nonumber \\ & - g_{ij} |V_j| \sin( \theta_j ) - (b_{ij}+b_{i0}) |V_i| \cos( \theta_i ) \nonumber \\ & - b_{ij} |V_j| \cos( \theta_j ) \label{eq:h_I2}. 
\end{align} 

Solving these non-linear equations is known to be a hard problem and often results in infeasible solutions \cite{AbuExp:04}. In general, three simplifying assumptions are made to reduce the AC model to a system of linear equations which can be efficiently solved - called the DC Power Model. The assumptions are: 1) $b_{ij} >> g_{ij}$; 2) Phase angle differences are negligible, i.e. $ \sin(\theta_i - \theta_j) \approx \theta_i - \theta_j$; 3) Voltage magnitudes are close to one, i.e. $|V_i| \approx 1$. Under the DC Model, then non-linearties collapse as:
\begin{align} \label{eq:DCmodel}
P_n = \sum_{i \neq n} b_{ni} (\theta_n - \theta_i)
\end{align}
The conjecture here is that, such simplifications can cause significant performance degradation for estimation algorithms. While the state-of-the-art algorithms as described in Section \ref{sec:soa} resort to linearizations of this sort, the main idea in this work is to numerically verify that indeed non-linearities in the system are non-neglectable and propose filtering methods that overcome this deficiency.

\subsection{State Estimation}
The problem of finding line voltage magnitudes and angels in the face of uncertainty can be cast as a \ac{HMM} (also referred to as Sequential Inference) problem. In this framework, based on the knowledge of past states, a new state is first predicted. This prediction is further corrected by taking into account the observables (i.e. measurements). This section details out this formulation and describes the methodology used in computing the different components.

First let's introduce some notation. Let
\begin{align}
\mathbf{x}_t= [  |V_1|,  |V_2| \cdots , |V_n|, \theta_1 , \theta_2, \cdots , \theta_n ]
\end{align}
denote the state vector consisting of voltage magnitudes and angles that we wish to estimate. As stated in Section \ref{sec:PowerSystem}, computing the state vector allows us to compute any other line values. That is, knowing the state vector tells us everything we need to know about the system. Simultaneously, let $\mathbf{y}_t$ be the vector of observations. To maintain generality, in this section let us avoid specifying what each element of the observation vector is. In general the entries of $\mathbf{y}_t$ can be any independent subset of the bus parameters at a given time $t$. In Section \ref{chap3}, for analysis this is explicitly set as:
\begin{align}
\mathbf{y}_t= [ |V_1|, |V_2| \cdots, |V_l| , P_1,  P_2, \cdots , P_m ].
\end{align}

The relation between state parameters and observations according to the discussion in Section \ref{sec:PowerSystem}, can ideally be expressed through the following relation: 
\begin{align} \label{eq:nonoiseobsmodel}
\mathbf{x}_t= h(\mathbf{y}_t)
\end{align}
To be specific, $h(\cdot)$ captures the observation equations (\ref{eq:h_P}-\ref{eq:h_Q}) pertinent to power measurements, and (\ref{eq:h_I1}-\ref{eq:h_I2}) pertinent to voltage and current measurements. This at first glance insinuates an algebraic-based solution via setting up a system of equations. However, there are three major considerations that call for a closer examination of the problem:

\begin{enumerate}[i)]
 \item The original relations between such parameters are highly non-linear.
 \item Even with linearization assumptions in (\ref{eq:DCmodel}), due to sparsity of measurements in large networks at any given time, not all state parameters can be obtained using this method.
 \item Observations themselves are inflicted by measurement noise.
\end{enumerate}
 
In view of the issues identified above, rather than a deterministic approach, it is then logical to take a probabilistic view - a common approach in the literature through Bayesian networks. In this framework, in lieu of direct computation of states, the joint probability density of the state vectors and observations $f(\mathbf{x}_{1:T}, \mathbf{y}_{1:T})$ from start to time $T$ is sought after. The joint probability density is of interest since other distributions of interest can be deduced via marginalization for instance, which in turn lend themselves to inference about single state parameters. The following sections explain the treatment of this density function.


\subsubsection{A Hidden Markov Model (HMM)} \label{sss:HMM}

Following the above discussion, the goal here is to make estimations about the state vector parameters based on noisy measurements of the observables. As an additional relation, the evolution of the states through time is also driven by system dynamics. For instance, the state of voltage on a given line is a function of powers drawn from connected branches. The specific system dynamics here can be described by a \ac{HMM} in which the hidden parameters are the voltages to be estimated and observables are any subset of voltage, power, and current measurements \cite{JorSej:01,Bis:06, ArnBalMinSar:08}.

\begin{figure}[t]
	\centering

	\psfrag{zk-1}[c][][1]{$y_{t-1}$}
	\psfrag{zk}[c][][1]{$y_t$}
	\psfrag{zk+1}[c][][1]{$y_{t+1}$}

	\psfrag{yk-1}[c][][1]{$x_{t-1}$}
	\psfrag{yk}[c][][1]{$x_t$}
	\psfrag{yk+1}[c][][1]{$x_{t+1}$}

	\psfrag{zk-1yk-1}[c][][1]{$f(y_{t-1}|x_{t-1})$}
	\psfrag{zkyk}[l][c][1]{$f(y_{t}|x_{t})$}
	\psfrag{zk+1yk+1}[c][][1]{$f(y_{t+1}|x_{t+1})$}
	
	\psfrag{ykyk-1}[Bc][][1]{$f(x_{t}|x_{t-1})$}
	\psfrag{yk+1yk}[Bc][][1]{$f(x_{t+1}|x_{t})$}
	
	\psfrag{ld}[c][][1]{$\cdots$}

	\ \\[0.06em]
	\hspace*{-3mm}
	\includegraphics[width=\columnwidth,keepaspectratio]{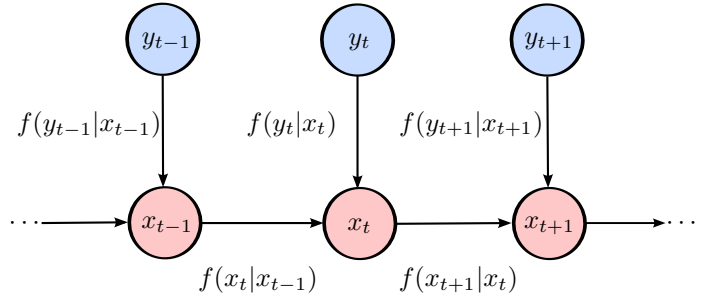}
   \caption{A \ac{HMM} for the sequential inference in Smart Grids.} 
   \label{fig:HMM}
\end{figure}

The \ac{HMM} framework leads to two immediate assumptions. Graphically, these conditions are expressed in Figure \ref{fig:HMM}:
\begin{enumerate}
 \item The states $ \mathbf{x}_t$ form a Markov chain, i.e. voltage magnitudes and angles at a time $t$ only depend on those of the previous and/or next time step.
 \item Observables $ \mathbf{y}_t$ are independent conditioned on the states $ \mathbf{x}_t$.
\end{enumerate}
The joint distribution of state and observation variables which is needed for the task of inference, under the \ac{HMM} assumptions, factors thereby as per (\ref{eq:jointpdffactorzation}):
\begin{align} \label{eq:jointpdffactorzation}
f(\mathbf{x}_{1:t}, \mathbf{y}_{1:t})  &=\prod_{i=1}^{t} f(\mathbf{x}_i | \mathbf{x}_{i-1}) f(\mathbf{y}_i|\mathbf{x}_i)  \nonumber \\  &= f(\mathbf{x}_{1:t-1}, \mathbf{y}_{1:t-1}) f(\mathbf{x}_t | \mathbf{x}_{t-1}) f(\mathbf{y}_t|\mathbf{x}_t)  
\end{align}

The state transition density function $f(\mathbf{x}_t | \mathbf{x}_{t-1})$ captures the Dynamic Model and the conditional observation likelihood density $f(\mathbf{y}_t|\mathbf{x}_t)$ captures the Observation Model, as shown on the graphical representation Figure \ref{fig:HMM}. In a physical system, the relation between states are captured by (\ref{eq:dynmodel}). Measurements and States are now related by a revised version of  (\ref{eq:nonoiseobsmodel}) and include a noise term for the uncertainty in the network given by:
\begin{align}
\mathbf{x}_{t+1} & = g( \mathbf{x}_t) + \mathbf{q}_t  \label{eq:dynmodel} \\
\mathbf{x}_t & = h(\mathbf{y}_t) + \mathbf{n}_t  \label{eq:obsmodel}
\end{align}
where $\mathbf{q}_t$ and $\mathbf{n}_t$ denote the dynamic model evolution and observation model uncertainty at time $t$, respectively. Equation (\ref{eq:dynmodel}) is referred to as the Dynamic Model and (\ref{eq:obsmodel}) as the Observation Model.

It is worth emphasizing an important implication of the factorization in (\ref{eq:jointpdffactorzation}). In general,  keeping track of the joint states and observables density $f(\mathbf{x}_{1:t},\mathbf{y}_{1:t})$ can be computationally cumbersome. However, the time series nature of the problem together with the structure of this \ac{HMM} allows at each time to compute $f(\mathbf{x}_{1:t} , \mathbf{y}_{1:t})$  recursively from the transition probability $f(\mathbf{x}_t | \mathbf{x}_{t-1})$ (i.e. Prediction Step) and the observation likelihood $f(\mathbf{y}_t | \mathbf{x}_{t})$ (i.e. Update Step). 

In view of this \ac{HMM} approach, we seek to compute the posterior density $f(\mathbf{x}_t | \mathbf{y}_{1:t})$ as an inference problem in a Bayesian Network. This computation through the multiple application of Bayes' rule is given by: 
\begin{align}
f(\mathbf{x}_{t} | \mathbf{y}_{1:t}) &= \frac{ f(\mathbf{y}_{t} | \mathbf{x}_{t} , \mathbf{y}_{1:t-1} ) f(\mathbf{x}_{t} | \mathbf{y}_{1:t-1})   }{ f(\mathbf{y}_{t} | \mathbf{y}_{1:t-1})  } \nonumber \\
 &= \frac{ f(\mathbf{y}_{t} | \mathbf{x}_{t}  ) \int f(\mathbf{x}_{t-1} | \mathbf{y}_{1:t-1}) f(\mathbf{x}_{t} | \mathbf{x}_{t-1} ) \mathrm{d}\mathbf{x}  }{ f(\mathbf{y}_{t} | \mathbf{y}_{1:t-1})  } 
\end{align}
This computation referred to as a filtering process, can be decomposed into two steps:  
\begin{align}
f(\mathbf{x}_{t}^{-} | \mathbf{y}_{1:t-1})  &\propto  \int f(\mathbf{x}_{t-1} | \mathbf{y}_{1:t-1}) f(\mathbf{x}_{t} | \mathbf{x}_{t-1} ) \mathrm{d}\mathbf{x}  \label{eq:Prediction}  \\
f(\mathbf{x}_{t} | \mathbf{y}_{1:t}) &\propto    f(\mathbf{y}_{t} | \mathbf{x}_{t}  ) f(\mathbf{x}_{t}^{-} | \mathbf{y}_{1:t-1})  \label{eq:Update}
\end{align}

Equation (\ref{eq:Prediction}) is referred to as the Prediction Step and (\ref{eq:Update}) is referred to as the Update Step. Our development in this work focuses on the latter. While the Prediction Step captures the dynamics of the system as it evolves through time, the implications of the assumptions in equations (\ref{eq:assumption2}-\ref{eq:assumption3}) are far too superior to neglect in achieving good performance. As such, we seek filtering methods that are immune to system non-linearities and complex noise models. In what follows, two such filters are presented - the \ac{BCF} and the \ac{PF}. Both methods have been applied in the context of Localization and Navigation Systems \cite{MazShiWin:13,WymLieWin:09} - an application with similar problem setup where positional state of agents and measurements are related through highly non-linear, non-gaussian relations. We present the details of the two filters below and later carry out analysis as to how much performance gain they achieve for the application of Smart Grids \ac{SE} in Section \ref{chap3}. The details of algorithmic implementation for the \ac{BCF} can also be found in \cite{mehryar2019belief}.

\subsubsection{Belief Condensation} \label{subsec:BC}

\ac{BCF} is a filtering framework where the true posterior of the state vector is approximated by a mixture of probability density functions. It has been shown that under certain optimality conditions, \ac{BCF} can provide accuracies approaching the theoretical bounds and outperforming existing techniques, particularly for non-linear/non-Gaussian problems. One of the main advantages of the \ac{BCF} filtering method is its treatment of the observation function as an inverse problem. Whereas Kalman-like filters tend to linearize the observation method where needed, which in specific cases of Gaussian distributions for instance perform remarkably well, the \ac{BCF} filter makes no assumptions as to what form $h(\cdot)$ must attain. This level of abstraction equipts the \ac{BCF} filter with interesting performance advantages while keeping the computational complexity low. In this section, the details of the \ac{BCF} filter are outlined, together with analysis that explores such advantages.

Consider the mixture family $\mathcal{F}_{\Xi^m}$ with an instance member $g(\mathbf{x}; \mathbf{\xi})$ given as:
\begin{align}
g(\mathbf{x}; \mathbf{\xi})= \sum_{i=1}^{m} \alpha_i g_i (\mathbf{x}; \mathbf{\theta}_i )
\end{align}
where $\{ \alpha_1, \alpha_2, \cdots, \alpha_m \} \in \mathbb{R}_+$, $\sum_{i=1}^{m} \alpha_i = 1$, and $ g_i (\mathbf{x}; \mathbf{\theta}_i )$, for each $i= \{ 1, 2, \cdots, m \}$, belongs to an exponential family $\mathcal{F}_{\Theta_m}$, given by:
\begin{align}
g_i(\mathbf{x}; \mathbf{\xi})= q_i(\mathbf{x}) \exp\{  \mathbf{\theta}_i^T \mathbf{t}_i(\mathbf{x}) - A_i(\mathbf{\theta_i}) \}
\end{align}
Here $\mathbf{\theta}_i \in \Theta_i$, $\mathbf{t}_i( \mathbf{x})$, and $A_i( \mathbf{\theta}_i )$ are the natural parameters, sufficient statistics, and log-partition function of $\mathcal{F}_{\Theta_i}$. The parameter set for $g(\mathbf{x}; \mathbf{\xi})$ consists of $ \mathbf{\xi}= ( \alpha_1, \mathbf{\theta}_1, \cdots, \alpha_m, \mathbf{\theta}_m) \in \Xi^m$.

Let $f \in \mathcal{P}$ from the distribution family $ \mathcal{P}$ denote the posterior distribution that we wish to approximate by $g(\mathbf{x}; \mathbf{\xi})  \in \mathcal{F}_{\Xi^m}$ within the Sequential Inference framework. For instance, in our particular analysis $f(\mathbf{x})=f( \mathbf{x}_{t} |  \mathbf{y}_{1:t} )$. The Kullback-Leibler (KL) divergence $D_{KL}$ between the probability distributions $f(\mathbf{x})$ and $g(\mathbf{x}; \mathbf{\xi})$  is defined by:
\begin{align}
D_{KL} \left ( f , g_{\mathbf{\xi}} \right ) = \mathbb{ E }_{f} \left \{ log \frac{f}{g_{\mathbf{\xi}} } \right \}
\end{align}
It can be shown that, under the following regularity conditions, \ac{BCF} recursions (see Theorem \ref{thrm1}), condense the probability distribution $f(\mathbf{x})$ into a mixture of exponential families:
\begin{description}
  \item[(A1)] The differential entropy of $f$ is finite
  \item[(A2)]  $\mathbb{ E }_{f} \{ | \log q_i(\mathbf{x})| \}$ is finite for each $i \in \{1,2, \cdots, m \}$
  \item[(A3)]  $\mathbb{ E }_{f} \{ | t_{i,j} ( \mathbf{x} ) | \}$ is finite, where $t_{i,j}(\mathbf{x})$ is the $j$-th component of the sufficient statistic $\mathbf{t}_{i}(\mathbf{x})$, for $i \in \{1,2, \cdots, m \}$
  \item[(A4)] The set $U_f = \{ \mathbf{\xi} \in \mathcal{F}_{\Xi^m} : D_{KL}( f, g_{  \mathbf{\xi} } ) < \infty \}$ is open and non-empty
\end{description}

\begin{theorem} \label{thrm1}
If a continuous probability distribution $f(\mathbf{x})$ satisfies the regularity conditions A1-A4, then the sequence $\{ D_{KL} ( f, g_{  \mathbf{\xi}^{[ l ]}   } ) \}_{l \in \mathbb{Z}_{+}}$ is monotonically decreasing, where the sequence $\{ \mathbf{\xi}^{[l]} \}_{l \in \mathbb{Z}_{+}} \subset \mathcal{F}_{\Xi^m}$ is recursively determined by 
\begin{align}
\alpha_i^{[l+1]} = \alpha_i^{[l]}  \mathbb{E}_{g_i \left ( \mathbf{x} ; \mathbf{\theta}_i^{[l]} \right ) } \left \{ \frac{f(\mathbf{x})}{g(\mathbf{x};\mathbf{\xi}^{[l]})} \right \}
\end{align}
for  $i=\{1,2,\cdots,m\}$ and $\mathbf{\theta}_i^{[l+1]}$ satisfying
\begin{align} \label{eq:recursivetheta}
 \mathbb{E}_{g_i \left ( \mathbf{x} ; \mathbf{\theta}_i^{[l+1]} \right )} \{ \mathbf{t}_i ( \mathbf{x} ) \} = \frac{  \mathbb{E}_{g_i \left ( \mathbf{x} ; \mathbf{\theta}_i^{[l]} \right )} \left \{ \frac{f(x)}{g(x; \mathbf{\xi}^{[l]})}  \mathbf{t}_i ( \mathbf{x} ) \right \}  }{  \mathbb{E}_{g_i \left ( \mathbf{x} ; \mathbf{\theta}_i^{[l]} \right )} \left \{ \frac{f(x)}{g(x; \mathbf{\xi}^{[l]})}  \right \} } 
\end{align}
for each $i=\{1,2,\cdots,m\}$ and any initial parameter $ \mathbf{\xi}^{[0]}= ( \alpha_1^{[0]}, \mathbf{\theta}_1^{[0]}, \alpha_2^{[0]}, \mathbf{\theta}_2^{[0]},  \cdots, \alpha_m^{[0]}, \mathbf{\theta}_m^{[0]}) \in U_f$.
\end{theorem}

In the case where the exponential families are Gaussian, $\mathbf{\theta}_i^{[l+1]}$ in (\ref{eq:recursivetheta}) can be obtained in a closed form as shown in the following Corollary. 

\begin{corollary}
Let $\mathcal{F}_{\Xi^m}$ be the mixture family of $m$ Gaussian distributions, with each mixture component parameterized by $\mathbf{\theta}_i=\{ \mathbf{\mu}_i , \mathbf{\Sigma}_i \}$. That is, 
\[
g_i (\mathbf{x}; \mathbf{\theta}_i ) \sim \mathcal{N} \left ( \mathbf{\mu}_i ,  \mathbf{\Sigma}_i \right )
\]
If $f(\mathbf{x})$ is a continuous probability distribution satisfying the regularity conditions A1-A4, then the natural parameter $\mathbf{\theta}_i^{[l+1]}$ at update step $l+1$ can be obtained by:
\begin{align}
 \mathbf{\mu}_i^{[l+1]} &= \frac{  \mathbb{E}_{g_i \left ( \mathbf{x} ; \mathbf{\theta}_i^{[l]} \right )} \left \{ \frac{f(x)}{g(x; \mathbf{\xi}^{[l]})} \mathbf{x}  \right \}  }{  \mathbb{E}_{g_i \left ( \mathbf{x} ; \mathbf{\theta}_i^{[l]} \right )} \left \{ \frac{f(x)}{g(x; \mathbf{\xi}^{[l]})}  \right \} } \label{eq:BCmuUpdate} \\
 \mathbf{\Sigma}_i^{[l+1]} &= \frac{  \mathbb{E}_{g_i \left ( \mathbf{x} ; \mathbf{\theta}_i^{[l]} \right )} \left \{ \frac{f(x)}{g(x; \mathbf{\xi}^{[l]})} \mathbf{x} \mathbf{x}^T  \right \}  }{  \mathbb{E}_{g_i \left ( \mathbf{x} ; \mathbf{\theta}_i^{[l]} \right )} \left \{ \frac{f(x)}{g(x; \mathbf{\xi}^{[l]})}  \right \} } \nonumber \\ &-  \mathbf{\mu}_i^{[l+1]} \left(  \mathbf{\mu}_i^{[l+1]}  \right)^T \label{eq:BCsigmaUpdate}
\end{align} 
\end{corollary}

Equations (\ref{eq:BCmuUpdate}) and  (\ref{eq:BCsigmaUpdate}) provide a recursive method for calculating and updating the state variables in each step. The main complexity in this computation comes from carrying out the computation for the expectation integrals. The fact that these expectations are taken with respect to a member of an exponential family (namely a Gaussian distribution) can be exploited, for which efficient quadrature rules exist \cite{StrSec:63,AraHayHur:10}. In this case, these integrals can be efficiently carried out with polynomial time in $m$ the number of components, in $q$ the number of quadrature points used, and in $d$ the dimension of the state vector.

\begin{example}[A Linear Model]
Consider a simple, linear observation model where we try to estimate parameter $x \in \mathbb{R}$ given $N$ i.i.d. noisy observations $y_i \in \mathbb{R}$, described by:
\[
y_i=x + n_i , \quad i=1, 2, \cdots , N
\]
where $n_i$ are independent, additive noise terms distributed according to a normal Gaussian distribution with variance $\sigma_n^2 = 0.1$. We run the \ac{BCF} filtering algorithm described above for different values of $m$, analyze, and compare the results to the \ac{EKF}, in order to provide insight about different aspects of the \ac{BCF} algorithm.

To characterize the performance of the \ac{BCF} algorithm, a Monte Carlo simulation is run and the \ac{MSE} over 100 trials is computed. In each trial, we generate $N=5$ noisy observations from the true value of $x=2.2$ and make point estimation using both \ac{BCF} and \ac{EKF} methods. We use the mean of the posterior as the point estimate. Figure \ref{fig:BCF_linear_MSEvsM_MC} shows the \ac{MSE} plot versus the size of gaussian mixture $m=\{1, 2, \cdots, 9\}$. Since the model is linear, the \ac{EKF} and \ac{BCF} show comparable performance (difference is within one standard deviation). As the number of gaussian mixtures used in the \ac{BCF} algorithm is increased, the accuracy also increases. After mixture size of $m=6$ however, the performance is degraded due to the fact that the model is becoming over-complex for the simple scenario under study here, hence impairing the accuracy. Figure \ref{fig:BCF_alpha_convrg} shows the convergence behavior of the coefficients $\alpha_i$ which in general settle after $l=5$ to $l=10$ steps.

\begin{figure}[t]
	\centering
	\psfrag{x}[tc][c][0.8]{Mixture size - $m$}
	\psfrag{y}[b][][0.8]{\ac{MSE}}
	\ \\[0.06em]
	\hspace*{-3mm}
	\includegraphics[width=\columnwidth,keepaspectratio]{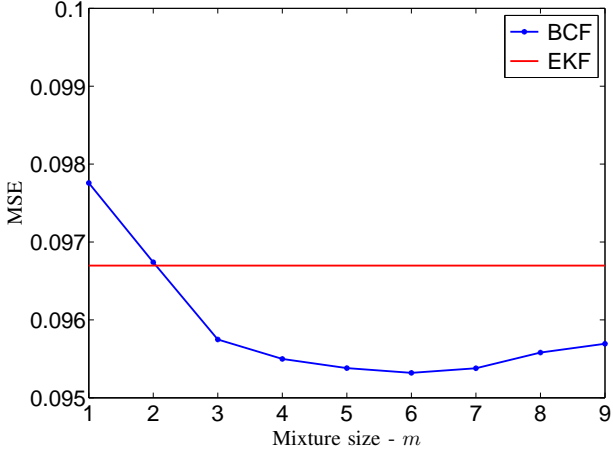}
   \caption{\ac{MSE} plot for a linear model using \ac{BCF} and \ac{EKF}.} 
   \label{fig:BCF_linear_MSEvsM_MC}
\end{figure}



\begin{figure}[t]
	\centering
	\psfrag{x}[tc][][0.8]{\ac{BCF} iterations - $l$}
	\psfrag{y}[b][][0.8]{$\alpha$ values}
	\ \\[0.06em]
	\hspace*{-3mm}
	\includegraphics[width=\columnwidth,keepaspectratio]{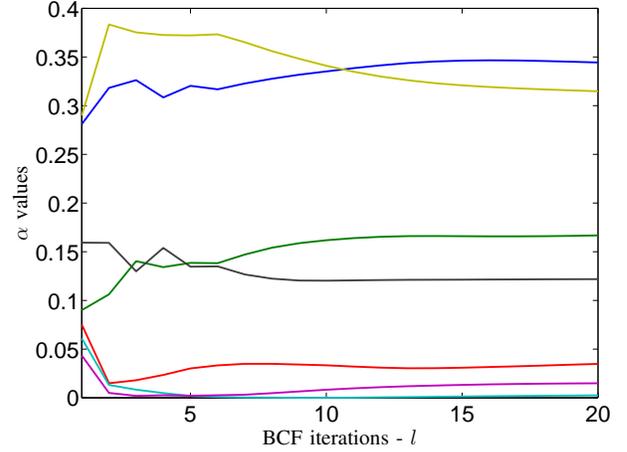}
   \caption{Convergence plots of $\alpha_i$'s in the \ac{BCF} algorithm.} 
   \label{fig:BCF_alpha_convrg}
\end{figure}

\end{example}

\begin{example}[A Bi-modal Distribution] \label{eg:ABimodalDistribution}
In order to further investigate the power of the \ac{BCF} filter in the face of more complex models, we consider a scenario where instead true values of $x$ are generated from a bi-modal distribution - a mixture of 2 gaussian distributions with means $\mu_1=-1.1$ and $\mu_2=1.1$, and variance $\sigma_n^2 = 0.1$. Here we set the mixing value to be $p=0.5$. Similar to the previous example, we run both the \ac{BCF} and \ac{EKF} algorithms and compare the behavior. In this case, we use $m=2$ gaussians in the mixture model. Figure \ref{fig:BCF_nongauss_pdf} shows that the \ac{BCF} filter represents this model correctly in the probability density sense, while the \ac{EKF} falls short in capturing the bi-modal nature of the underlying model and is concentrated around the mean with small standard deviation.

%

\begin{figure}[t]
	\centering
	\psfrag{x}[tc][][0.8]{$x$}
	\psfrag{y}[b][][0.8]{Probability density function}
	\ \\[0.06em]
	\hspace*{-3mm}
	\includegraphics[width=\columnwidth,keepaspectratio]{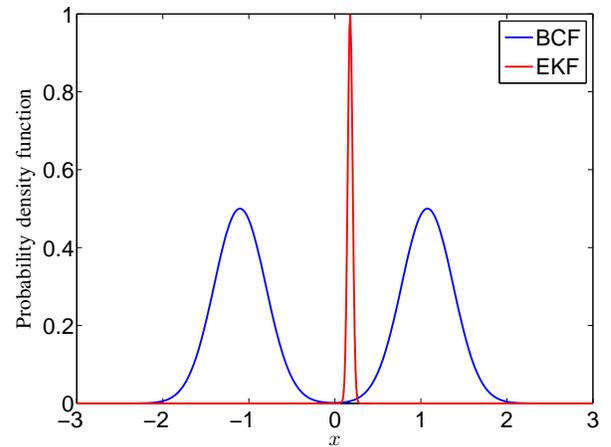}
   \caption{Probability density functions for a bi-modal parameter as computed by the \ac{BCF} and \ac{EKF}.} 
   \label{fig:BCF_nongauss_pdf}
\end{figure}

\end{example}

\begin{example}[A Non-linear Model]
As a final example, let's consider an observation function that has a general form similar to those that appear in Power Grids. In particular, imagine the relation between the observation $y$ and state parameter $x$ to be expressed by the non-linear function $h(x)=x^2 \sin(x)$, which gives rise to the following observation model:
\[
y_i = x^2 \sin(x) + n_i , \quad i= 1,2, \cdots N
\]
As before, we run 100 Monte Carlo trials and compute the \ac{MSE} from both the \ac{BCF} and \ac{EKF} algorithms. The results are shown in Figure \ref{fig:BCF_nonlinear_MSEvsM}. From the figure, it can be seen that in the presence of non-linearity in the model, the \ac{BCF} outperforms the Kalman filter due to linearization assumptions such methods make. It can also be seen that, with increasing $m$, the accuracy in estimation for \ac{BCF} also improves.


\begin{figure}[t]
	\centering
	\psfrag{x}[tc][][0.8]{Mixture size - $m$}
	\psfrag{y}[b][][0.8]{\ac{MSE}}
	\ \\[0.06em]
	\hspace*{-3mm}
	\includegraphics[width=\columnwidth,keepaspectratio]{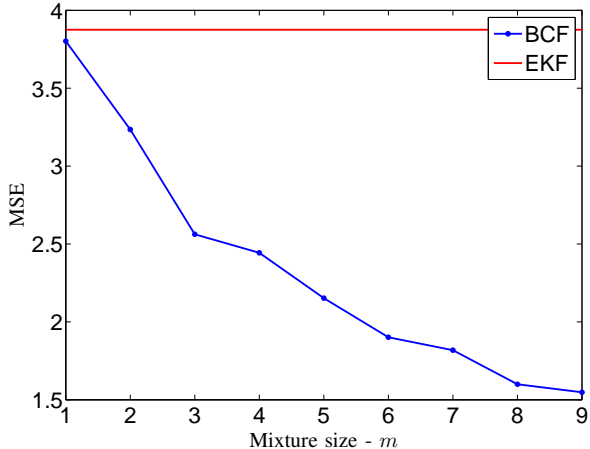}
   \caption{\ac{MSE} plot for a non-linear model using \ac{BCF} and \ac{EKF}.} 
   \label{fig:BCF_nonlinear_MSEvsM}
\end{figure}

\end{example}

\subsubsection{Particle Filtering} \label{subsec:PF}

A density representation method that has recently obtained much popularity in a variety of applications is the \ac{PF}. These filters are a powerful tool that theoretically can approximate any density function in the limit of sampling size. In essence, this method discretizes the probably space via an ensemble of samples called particles, each of which is weighted to represent the density at that sample. Therefore, as the number of samples $n$ approaches infinity, the gap between the true and approximated distribution closes ever so tightly. In this section, the mathematical formulation of the \ac{PF} method is explained, along with examples to provide insightful perspective.

The \ac{PF} representation of a posterior density of interest $f(\mathbf{x})$ consists of a set $\{\hat{ \mathbf{x}}_i , w_i \}_{i=1}^n$ of $n$ particles (or samples) $\hat{\mathbf{x}}_i$ and their respective weights $w_i$, for each $i \in \{1, 2, \cdots, n \} $.  Consequently, we can write an approximate representation for $f(\mathbf{x})$ with: 
\begin{align}
f(\mathbf{x}) \approx \sum_{i=1}^{n} w_i \delta ( \mathbf{x}   ,\hat{ \mathbf{x} }_i   ) \label{eq:PCmuUpdate}
\end{align}
where $\delta ( .)$ is the Dirac delta function, and weights sum to one,  $\sum_{i=1}^{n} w_i =1  $. As $n \to \infty$, the approximation would be exact in theory, i.e. $f(\mathbf{x}) = \sum_{i=1}^{\infty} w_i \delta ( \mathbf{x}   ,\hat{ \mathbf{x} }_i   )   $.

There exist a number of different implementations of the \ac{PF}, a survey of which is provided in \cite{PFsurvey,GorSalSmi:93,MerDouFreWan:00}. In this work, given the particle representation for a density function $f(\mathbf{x})$, the recursive update equation for sample weights is given by \cite{AruMasGorCla:02}: 
\begin{align}
w_{i}^{[l+1]} = f(\mathbf{y}^{[l]} | \hat{\mathbf{x}}_{i}^{[l]} )  w_{i}^{[l]}
\end{align}
and the task of inference on the state vector $x$ can be then carried out via the \ac{MMSE} given by :
\begin{align}
\hat{\mathbf{x}}_\text{MMSE} = \sum_{i=1}^{n} w_i \hat{ \mathbf{x}}_i   
\end{align}
or the Maximum A-Posteriori (MAP) estimate given by:   
\begin{align}
\hat{\mathbf{x}}_\text{MAP} = \argmax_{\hat{ \mathbf{x}}_1 , \cdots,  \hat{ \mathbf{x} }_n    } \{ w_i \}
\end{align}

\begin{example}[A Bi-modal Distribution] \label{eg:ABimodalDistributionPF}
Let us revisit Example \ref{eg:ABimodalDistribution} where previously a \ac{BCF} filtering method was used to represent the true posterior density. Recall, in that example the true values of $x$ are generated via a bi-modal distribution $p \mathcal{N}(-1.1,0.1) + (1-p)  \mathcal{N}(1.1,0.1)$, with $p=0.5$. Here instead $n=1000$ particles is used to represent the true distribution. Figure \ref{fig:PF_nongauss_pdf} shows the result of such an approximation. It can be seen that, much like the \ac{BCF}, the \ac{PF} is able to represent complex distributions, albeit through discretized samples.


\begin{figure}[t]
	\centering
	\psfrag{x}[tc][][0.8]{$x$}
	\psfrag{y}[b][][0.8]{Probability density function}
	\ \\[0.06em]
	\hspace*{-3mm}
	\includegraphics[width=\columnwidth,keepaspectratio]{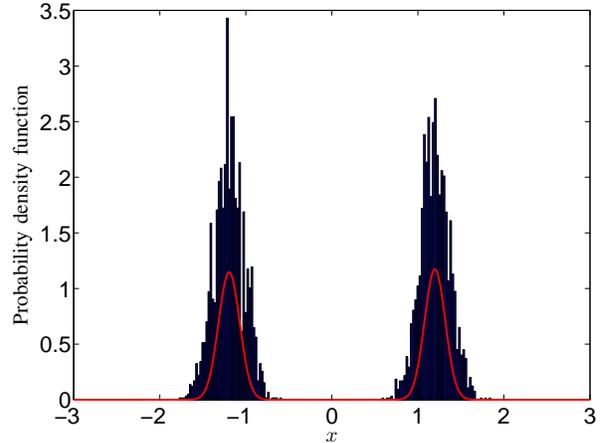}
   \caption{Probability density functions for a bi-modal parameter as computed by the \ac{PF}. The red curve shows a gaussian fit to what are weighted particles shown as blue bars.} 
   \label{fig:PF_nongauss_pdf}
\end{figure}

\end{example}

\section{Numerical Experimentation and Analysis} \label{chap3}

In this section, we apply the \ac{BCF} and \ac{PF} methodologies to the problem of \ac{SE} in power networks. The main goal of this experimentation is to investigate and numerically support the following claims:
\begin{enumerate}[i)]
  \item In the presence of non-linearities in the system, Kalman-like filters that resort to linear approximations, trade-off accuracy for efficiency. In such cases, continuous mixture models (e.g. \ac{BCF}) and discretized alternatives (e.g. \ac{PF}) provide an ideal substitute, maintaining both accuracy and efficiency.
  \item In the presence of more complex and/or unknown noise (e.g. non-Gaussian) in the system, the said methods (i.e. \ac{BCF} and \ac{PF}) significantly reduce the estimation error by abstracting out the noise model.
\end{enumerate}

In order to quantify the outcomes under each claim, first a measure of accuracy and a measure of efficiency are established. Characterizing the accuracy of the algorithms is done via the \ac{MSE} per unit \footnote{All variables are normalized to avoid unit mismatches in \ac{MSE} computation. This is due to the fact that the state vector contains both voltage and angle values which have different units.}. Given estimates $\hat{ \mathbf{x}}_t$ at time $t$, the \ac{MSE} is defined by:
\begin{align}
\text{MSE} = \frac{1}{T} \sum_{t=1}^{T} ||\mathbf{x}_{\text{true}}-\hat{ \mathbf{x}}_t||^2
\end{align}
where $\mathbf{x}_{\text{true}}$ denotes the true values of the state vector, and the mixture mean and the \ac{MMSE} are used as the final point estimate in \ac{BCF} and \ac{PF}, respectively. The efficiency is analyzed by looking at either the computation complexity using the $\mathcal{O}(\cdot)$ notation or the computation time per operation, depending on the type of algorithm under study.

Two popular methods from the literature are used to benchmark the results: 1) \ac{LSQ} - a generic method that is widely used in the current power networks; 2) \ac{UKF} - the most state-of-the-art Kalman-like filter in the community. 

Using the measures defined here, the claims made above are verified through $3$ sets of numerical experiments. In the first set of simulations described in Section \ref{sec:LandNLF}, the effect of linear (Kalman) versus non-linear (\ac{BCF} and \ac{PF}) filleting is investigated by computing the \ac{MSE} of each algorithm. By holding the dynamic changes and noise model constant in the system, in this case the performance behaviors of the \ac{BCF} and \ac{PF} are singled out and measured against the \ac{UKF}.

In the second set of simulations, in Section \ref{sec:Noise} a mixture noise model is introduced that is additive, identically distributed, and is controlled for Gaussian-ness by a parameter $p$ - the noise mixing coefficient. In a similar fashion, the resiliency of each algorithm to noise is measured in terms of the \ac{MSE} as a function of varying noise coefficient. Lastly, in Section \ref{sec:FullSim}, the filtering methods are simulated in a real-time setting where the state values are dynamically changed over time. In this important step, the goal is to address issues related to practical and operational aspects of the proposed methods in a real life power grid scenario. In particular, we are interested in knowing the performance gain in employing non-linear filters as an alternative in future power grids.

\subsection{Simulation Set-up} \label{sec:SimSetup}

\begin{figure}[t]
  \centering
  \includegraphics[width=0.45\textwidth,height=0.95\textwidth,keepaspectratio]{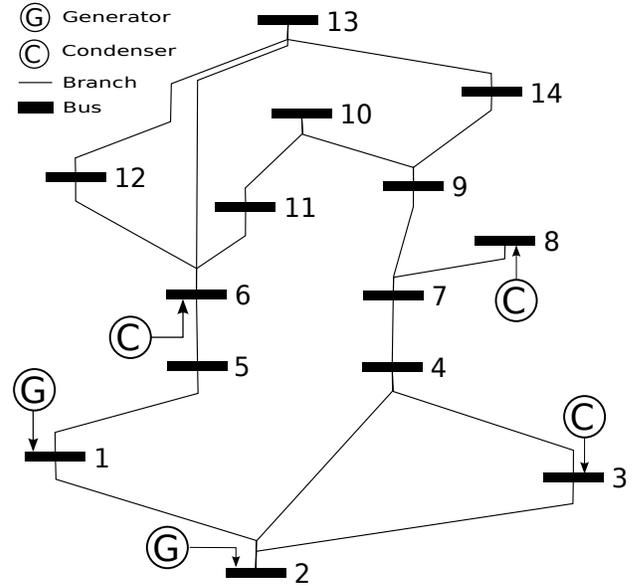}
  \caption{IEEE 14-bus test system.}
  \label{fig:14bus}
\end{figure}

In order to verify the performance improvements by the proposed methods in this work, the IEEE 14-bus test system is utilized for numerical simulation. This test system is available for free access at \cite{ZimMurTho:11} as a part of the MATPOWER package which contains full line state variables and parameters. Figure \ref{fig:14bus} shows a diagram of the IEEE 14-bus system. Bus $1$ is assigned as the slack/reference bus ($|V_1|=1$ and $\theta_1=0$). The task is to estimate $\{|V_2|, |V_3|, \cdots, |V_{14}|, \theta_2, \theta_3, \cdots, \theta_{14}\}$ corresponding to Bus $2$ through to $14$. In all the experiments that follow, a single iteration of the \ac{BCF} and \ac{PF} ($l=1$) is performed in each trial. 

Measurements used in the Observation Model are taken at random from power injections as well as voltage amplitude, for a total of $28$ measurements in each time step. The measurements are generated by running the real state values through the observation function and superimposing an additive noise. The Gaussian portion of all noise models hereon have a zero mean with standard deviation $2\%$ and $0.1\%$ for power and voltage measurements, respectively. The mixture noise models are described as necessary in the following sections.

In order to simulate the slow dynamics of the systems, a subset of loads is selected initially and changed over a period of 50 time sample intervals. The loads are varied following a linear trend of increase or decrease by $10\%$ in each time step. In addition to these linear changes, a fluctuation of $3\%$ is also superimposed. The dynamic model used in the simulations is the Holt's Smoothing method, with elements of diagonal matrix $\mathbf{Q}$ equal to $10^{-6}$ and $\mathbf{P}_0$ initially equal to $10^{-6}$. Since Holt's method is 2-step smoothing filter, it's assumed that the state of the lines are known at first, and the simulation starts at $t=2$. 


\subsection{Linear and Non-Linear Filtering} \label{sec:LandNLF}

As the first case in point, let us investigate the performance characteristics in the face of non-linearities in the system model for the \ac{BCF} and \ac{PF} filtering methods and their treatment thereof. As argued in the previous sections, non-linearities arise in the system due to the complex nature of power equations captured by the Observation Model. In order to isolate this effect, let us fix the time-varying dynamics and consider the network in a static mode. This reduces the network equations down to a single Observation Model given by:
\begin{align}
\mathbf{x}_t= h(\mathbf{y}_t) + \mathbf{n}_t  \quad 
\end{align}

Each method is tested by estimating $h(\mathbf{y}_t)$ through Monte Carlo sampling from $1000$ independent measurements $\mathbf{y}_t$, from which an aggregate \ac{MSE} is computed. This is done for different settings of \ac{BCF} with $m=1, 2, \cdots, 9 $ components and \ac{PF} with $n=500$ to $n=10000$ particles. The noise here is assumed to be i.i.d. Gaussian. The fundamental question here is, under equal dynamic and noise models, do non-linear filters achieve better accuracy (\ac{MSE}) and efficiency than linear filters?

\begin{figure}[t]
  \centering
  
        \psfrag{x}[tc][][0.8]{Number of components - $m$}
	\psfrag{y}[b][][0.8]{\ac{MSE} per unit per line}
	\ \\[0.06em]
	\hspace*{-3mm}
	\includegraphics[width=\columnwidth,keepaspectratio]{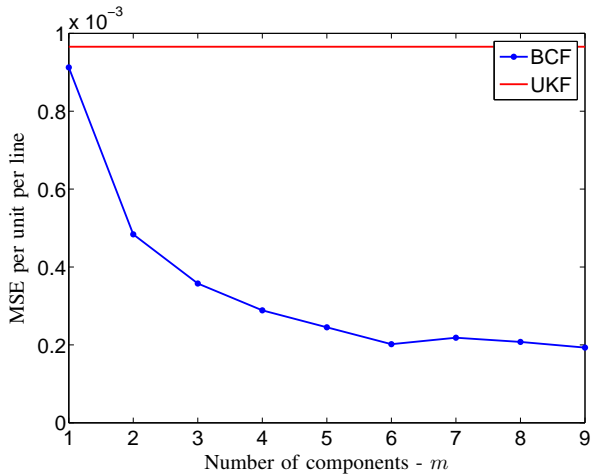}

  \caption{\ac{MSE} computation for the \ac{BCF} as a function of the number of mixture components against the \ac{UKF}.}
  \label{fig:BC_m}
\end{figure}

Figure \ref{fig:BC_m} shows the \ac{MSE} as a function of mixture components $m$ for \ac{BCF}. Inset the figure also is \ac{UKF}'s \ac{MSE} - a constant line in this case. It can be seen that in general the \ac{BCF} curve lies below the \ac{UKF} line, suggesting performance improvements by non-linear filtering. 

From the figure, at $m=1$ it can be seen that \ac{BCF} achieves a slight improvement in accuracy. This gain in performance is due to \ac{BCF}'s treatment of non-linear equations $h(\cdot)$ as an inverse problem. Based on the discussion in Section \ref{subsec:BC}, while Kalman-like filtering methods resort to linearization, the \ac{BCF} estimates the density function for the state variables from the measurements using the true observation model. With more components, i.e. increasing $m$, the complexity of the \ac{BCF} model increases, resulting in better estimation power. It can be seen that at $m=9$, near an order of magnitude increase in accuracy can be obtained by non-linear filtering (the \ac{BCF}) versus linear filtering (the \ac{UKF}). Ultimately, with a minimum of $m=6$ gaussian components the \ac{BCF} performance is optimized.

To give this discussion context, let us revisit the IEEE 14-bus test system and re-interpret the results. For a sub-network that contains $14$ bus lines at nominal voltage levels of $110V$ per line (typical in U.S. households), the \ac{UKF} and the baseline \ac{BCF} ($m=1$) have net \ac{MSE}'s of approximately $1.5V$ and $1.3V$, respectively \footnote{The net \ac{MSE} is computed as the value of \ac{MSE} per unit per line $\times$ the nominal voltage value $\times$ the number of bus lines .}. \ac{BCF} with $m=2$ components almost doubles the performance with the net \ac{MSE} of $0.75V$. With $m=6$ components, the net \ac{MSE} is reduced even further to $0.31V$.

\begin{figure}[t]
  \centering
  
        \psfrag{x}[tc][][0.8]{Number of particles - $n$}
	\psfrag{y}[b][][0.8]{\ac{MSE} per unit per line}
	\ \\[0.06em]
	\hspace*{-3mm}
	\includegraphics[width=\columnwidth,keepaspectratio]{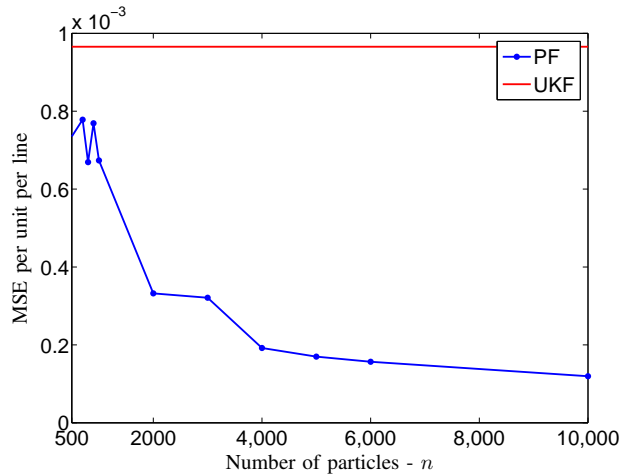}

  \caption{\ac{MSE} computation for the \ac{PF} as a function of the number of particles compared against the \ac{UKF}.}
  \label{fig:PF_n}
\end{figure}

Figure \ref{fig:PF_n} shows the \ac{MSE} curve for the \ac{PF}, versus \ac{UKF}, as a function of the number of particles. For the number of particles $n$ under 1000, we can see an oscillatory behavior in the \ac{PF} algorithm. This is due to insufficient particles resulting in unstable estimation. From the figure, at about $n=4000$ particles the best accuracy (i.e. lowest \ac{MSE}) achieved by $m=9$ components from \ac{BCF} is surpassed. At $n=10000$ particles, the \ac{PF} obtains an order of magnitude improved performance over the \ac{UKF}. For the IEEE 14-bus test system, this corresponds to a net \ac{MSE} of $18 mV$.

\subsection{Non-gaussian Noise Model} \label{sec:Noise}


In the previous section, it became evident that in the presence of non-linearities, filters that abstract out the role of the observation function are needed in order to achieve better accuracy. Other elements of the model were held constant to verify this conjecture. Specifically, the observation noise was treated as additive, independent, and gaussian with:
\begin{align}
\mathbf{n}_t \sim \mathcal{N}(\mathbf{\mu}_n^t, \mathbf{\Sigma}_n^t)
\end{align}  
While this noise model suffices in cases where only device-related measurement errors are of interest, in more realistic networks the true noise model is not known. For instance, with the advent of wireless \ac{PMU} and meters, communication errors due to packet drops among others may contribute to measurement errors. In such scenarios, more sophisticated filtering methods are required that can alleviate uncertainties in the system. 

In this section, let's investigate the effect of non-gaussian error in power grids and expose the filtering methods to more general error models to understand their resilience. In particular, let us consider a general mixture noise model given by:
\begin{align}
\mathbf{n}_t = p_e \mathbf{n}_t^g  + (1-p_e) \mathbf{n}_t^u
\end{align}
which is the sum of a gaussian random variable $\mathbf{n}_t^g$ and a uniform random variable $ \mathbf{n}_t^u$, regulated by the mixture coefficient $p_e \in [0,1]$: 
\begin{align}
 \mathbf{n}_t^g  &\sim \mathcal{N}(\mathbf{\mu}_n^t  \mathbf{\Sigma}_n^t) \\
\mathbf{n}_t^u &\sim \mathcal{U}(a,b)
\end{align}

The uniform component can be thought of as an event where the measurement is fully corrupted due to packet collision for instance, with probability of occurrence $1-p_e$. Let's repeat the simulation process as above, and look at the \ac{MSE} as a function of the mixing coefficient $p_e$. Notice that at $p_e=1$, the numerical results collapse to that seen in Section \ref{sec:LandNLF}, while at the other extreme when $p_e=0$, we only see uniform noise. 

\begin{figure}[t]
  \centering
  
        \psfrag{x}[tc][][0.8]{Noise mixing coefficient - $p_e$}
	\psfrag{y}[b][][0.8]{\ac{MSE} per unit per line}
	\ \\[0.06em]
	\hspace*{-3mm}
	\includegraphics[width=\columnwidth,keepaspectratio]{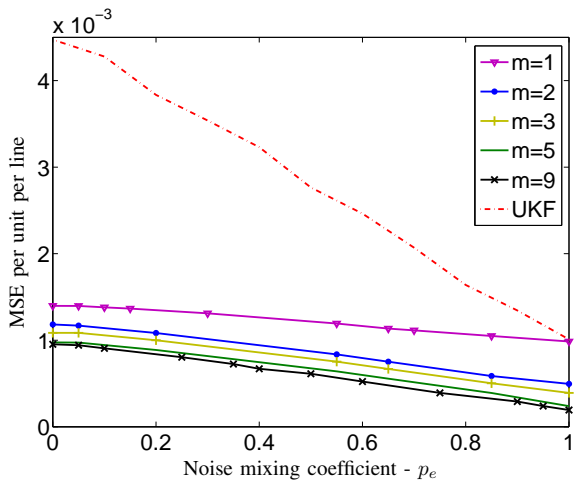}

  \caption{\ac{MSE} for \ac{BCF} against \ac{UKF}, as a function of mixing coefficient $1-p$. At the extreme points $p=1$ and $p=0$, the noise is fully non-gaussian and fully gaussian, respectively.}
  \label{fig:BC_p}
\end{figure}

Figure \ref{fig:BC_p} shows the effect of non-gaussian noise on the \ac{BCF} algorithm with $m=1, 2, 3, 5, 9$ components in terms of the \ac{MSE}. The \ac{UKF} curve is shown in red. In general, there is an increasing pattern as the noise mixture approaches uniform distribution (to the left). This is expected since uniform distribution implies the most uncertainty in the system, contributing most to the degradation of the estimates. The main point to note from this figure is the relative degradation in the \ac{BCF} as opposed to the \ac{UKF}. While due to increasing uncertainty the \ac{UKF}'s \ac{MSE} experiences a $4.4$-fold increase, the \ac{BCF} exhibits more noise resilience, keeping this degradation gradual and limited to under $2$ times. As expected, the \ac{BCF}'s performance improves with the number of components included (i.e. $m=2, 3, 5, 9$) due to better accuracy in the update step through the observation model.

\begin{figure}[t]
  \centering
  
         \psfrag{x}[tc][][0.8]{Noise mixing coefficient - $p_e$}
	\psfrag{y}[b][][0.8]{\ac{MSE} per unit per line}
	\ \\[0.06em]
	\hspace*{-3mm}
	\includegraphics[width=\columnwidth,keepaspectratio]{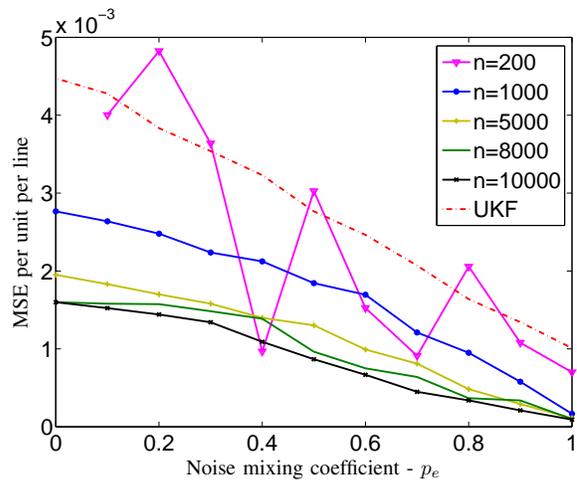}

  \caption{\ac{MSE} for \ac{PF} against \ac{UKF}, as a function of mixing coefficient $1-p$. At the extreme points $p=1$ and $p=0$, the noise is fully non-gaussian and fully gaussian, respectively.}
  \label{fig:PF_p}
\end{figure}

Figure \ref{fig:PF_p} shows these results for the \ac{PF} with $n= 200, 1000, 5000, 8000, 10000$ particles. At very low number of particles, the \ac{PF} exhibits oscillatory behavior, seen from the curve for $n=200$. This is due to insufficient number of particles to rightfully represent the true distribution (under $1000$ particles according to discussion in Figure \ref{fig:PF_n} Section \ref{sec:LandNLF}). With larger number of particles, the performance also improves. As was the case with \ac{BCF}, with increasing uncertainty (from right to left on the figure) the performance is impaired. In contrast to the \ac{BCF}, at smaller number of particles, the \ac{PF} has a steeper increasing pattern with the mixing noise coefficient. It can be seen that with the number of particles $n$ large enough, the steadiness is regained. At $n=10000$, the change in \ac{MSE} is still 16-fold - a considerable jump. This is attributed to the fact that the \ac{MMSE} estimate is a weighted sum of all the particles that are overfitting to the complex noise model in this case.

\subsection{Full System Simulation}  \label{sec:FullSim}

 
As a final test, in this section we put the pieces together and a full simulation is carried to compare algorithms as the state changes through time. The evolution of states through time is described by the Dynamic Model given in (\ref{eq:dynmodel}). To this end, a subset of the state variables are selected pre-run at random. During the simulations, the selected subset's true values are changed according to a linear model (increased or decreased) as per descriptions in Section \ref{sec:SimSetup}. Measurement noise is considered in the same fashion as before.

In each time step, first a prediction is made based on the previously computed state values and measured observations. This step corresponds to Prediction Step in (\ref{eq:Prediction}) in the \ac{HMM} model. Here, Holt's smoothing is used that provides a linearized way (See Appendix \ref{Appx:Holtz} for details). In Holt's method, innate parameters are set as $\alpha_t=0.81$ and $\beta_t=0.56$, empirically. Once a new state is obtained, predictions are updated with measurements using each of the algorithms under study, in a set-up similar to previous sections.

\begin{figure}[t]
  \centering

         \psfrag{x}[tc][][0.8]{Time step - $t$}
	\psfrag{y}[b][][0.8]{\ac{MSE} per unit}
	\ \\[0.06em]
	\hspace*{-3mm}
	\includegraphics[width=\columnwidth,keepaspectratio]{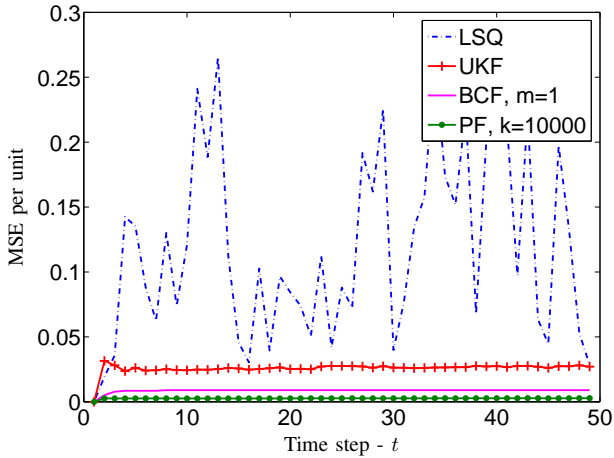}
  
  \caption{\ac{MSE} for four different filtering algorithms: 1) \ac{LSQ}, 2) \ac{UKF}, 3) \ac{BCF}, and 4) \ac{PF}.}
  \label{fig:MSE_t}
\end{figure}

Figure \ref{fig:MSE_t} shows the \ac{MSE} per unit for 4 different filtering methods - \ac{LSQ}, \ac{UKF}, baseline \ac{BCF} with $m=1$, and the \ac{PF} with $n=10000$. The \ac{LSQ} filter as a fourth method is considered here due to wide acceptance currently in the community as an estimation filter. In each time step, a prediction $\mathbf{x}_t^{-}$ is made for the state vector according to the dynamic model (same for all filtering methods), followed by an update step by each filter to obtain  $\mathbf{x}_t$.

It can be seen that the baseline \ac{BCF} and \ac{PF} reduce the \ac{MSE} by an order of magnitude per unit over their counter parts. The \ac{LSQ} has the poorest and most fluctuating performance. This is due to the fact that it minimizes the squared error between the measurements $\mathbf{y}_t$ and $h(\mathbf{x}_t)$ using a gradient descent approach. It disregards non-linear effects and lacks a smoothing mechanism over time. The remaining curves exhibit a smooth prediction and correction tracking overall. Due to their strength in mitigating the effect of non-linearities, the \ac{BCF} and \ac{PF} curves lie below the \ac{UKF} curve. The \ac{UKF} suffers from the highly non-linear nature of the observation model, resulting in impaired performance as expected.  

\begin{figure}[t]
	\centering
	\psfrag{x}[tc][][0.8]{Time step - $t$}
	\psfrag{y}[b][][0.8]{\ac{MSE} per unit}
	\ \\[0.06em]
	\hspace*{-3mm}
	\includegraphics[width=\columnwidth,keepaspectratio]{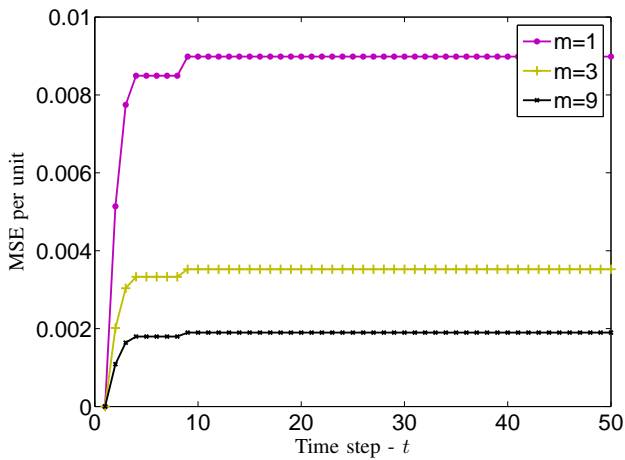}
   \caption{\ac{MSE} for the \ac{BCF} algorithm with $m=1,3,9$ components.} 
   \label{fig:BCdyno}
\end{figure}

To further investigate the behavior of \ac{BCF} algorithm in a dynamic setting, the same experiment is repeated for $m=3$ and $m=9$. The results are shown in Figure \ref{fig:BCdyno} along the $m=1$ plot from Figure \ref{fig:MSE_t}. As before, the there filters start and settle into a steady state tracking the linear changes in the network. The slight jump in the plots at time step $10$ is where the state of randomly selected bus lines start to change for all simulations, so to leave Holt's method slack to settle in. As more \ac{BCF} components are added to \ac{BCF}, the update from the measurements is improved in each step. These incremental improvements carry over to next time steps through the dynamic model, making the overall performance better as confirmed by $m=3$ and $m=9$ curves. Furthermore, in Section \ref{sec:LandNLF} it was observed that with enough components, the \ac{BCF} can achieve accuracies close to the \ac{PF}. From Figure \ref{fig:BCdyno}, with $m=9$ the dynamic \ac{BCF} achieves accuracy of around $0.002$ per unit or $0.29$ Volts. This is almost identical to $PF$'s performance at $n=10000$ particles which confirms our observation.

Average \ac{MSE} values over $500$ runs are shown in Table \ref{table:MSE}. To put these numbers in perspective, let us revisit the IEEE 14-bus example that is used as a test bench in this experimentation. Reading from the third column in Table \ref{table:MSE}, in this test case (14 buses, with each bus at nominal voltage $110V$) the \ac{LSQ} has an \ac{MSE} of $14 V$. For a relatively small sized network such as the IEEE 14-bus, this is a considerable error in estimation. The \ac{UKF} has an \ac{MSE} of $2.9 V$, a much smaller deviation comparatively. The \ac{BCF} further keeps the error margin under a volt, while the \ac{PF} algorithm reduces \ac{MSE} to a few hundred milli-volts. This is an improvement in accuracy by at least an order magnitude.  

\begin{table}
\caption{\ac{MSE}.}
\label{table:MSE}
\begin{center}
\rowcolors{2}{lightgray}{}
\begin{tabular}{c|c|c|}
\cline{2-3}
 & \textbf{\ac{MSE} (per unit)} & \textbf{\ac{MSE} (IEEE 14-bus at $\mathbf{110 V}$)}  \\\hline
\multicolumn{1}{|c|}{\ac{LSQ}}	   & $0.1275$ & $14 V$ \\
\multicolumn{1}{|c|}{\ac{UKF}}	   & $0.0264$ & $2.9 V$ \\
\multicolumn{1}{|c|}{\ac{BCF}}	   & $0.0089$ & $0.97 V$ \\
\multicolumn{1}{|c|}{\ac{PF}}	   & $0.0026$ & $0.29 V$ \\\hline
\end{tabular}
\end{center}
\end{table}

\subsection{Complexity Analysis} \label{Appx:cplx}

The results thus far have shown that non-linear filters such as the \ac{BCF} and \ac{PF} improve upon Kalman-like filters in estimating the state of power grids. In particular, the \ac{PF}, with the right ensemble size, outperforms the rest of the methods (see Table \ref{table:MSE}). In the current literature, it is claimed that in the limit this filtering method indeed can achieve the theoretical bounds of estimation accuracy. On the other hand, it is known that \ac{PF} suffers from the curse of dimensionality as it utilizes Monte Carlo methods to approximate multidimensional integrals \cite{AruRis:00, MusGreKasKre:01, FarRisBen:02}. To this date, there is no analytical formula that capture the complexity of \ac{PF} as a function of the size and computation time of the problem.

In light of this obstacle and in order to trade off accuracy versus practical considerations, we pursue and compare here the computation time of the \ac{BCF} and BF through numerical experimentation. We adopt a methodology presented in \cite{DauHua:03}, whereby a dimension free metric for error is introduced. The dimension-less nature of this metric allows us to abstract out performance degeneration caused due the size of the problem and get a true comparison by fixing the accuracy.

Adopting the notation of \cite{DauHua:03}, let $r$ denote the Mean Dimension-Free Error defined by:
\begin{align}
r = \frac{ \mathbb{ E }\left \{(x-y)^{*} J (x-y)  \right \} }{d}
\end{align}
where $y$ is the estimate of $x$ from the \ac{BCF} or \ac{PF}, $J$ is the inverse of the estimation error covariance matrix, $x$ is the state vector to be estimated, and $(\cdot)^*$ denotes the transpose of $(\cdot)$. First the number of mixture components $m$ (in the case of \ac{BCF}) or particles $n$ (in the case of BF) are selected that obtain a fixed value of $r$. The complexity of each filter is then defined as the time it takes the algorithms to reach $r$ with parameters selected in this fashion.

\begin{figure}[t]
	\centering
	\psfrag{x}[tc][][0.8]{Number of components - $m$}
	\psfrag{y}[b][][0.8]{Mean dimension-free error - $r$}
	\ \\[0.06em]
	\hspace*{-3mm}
	\includegraphics[width=\columnwidth,keepaspectratio]{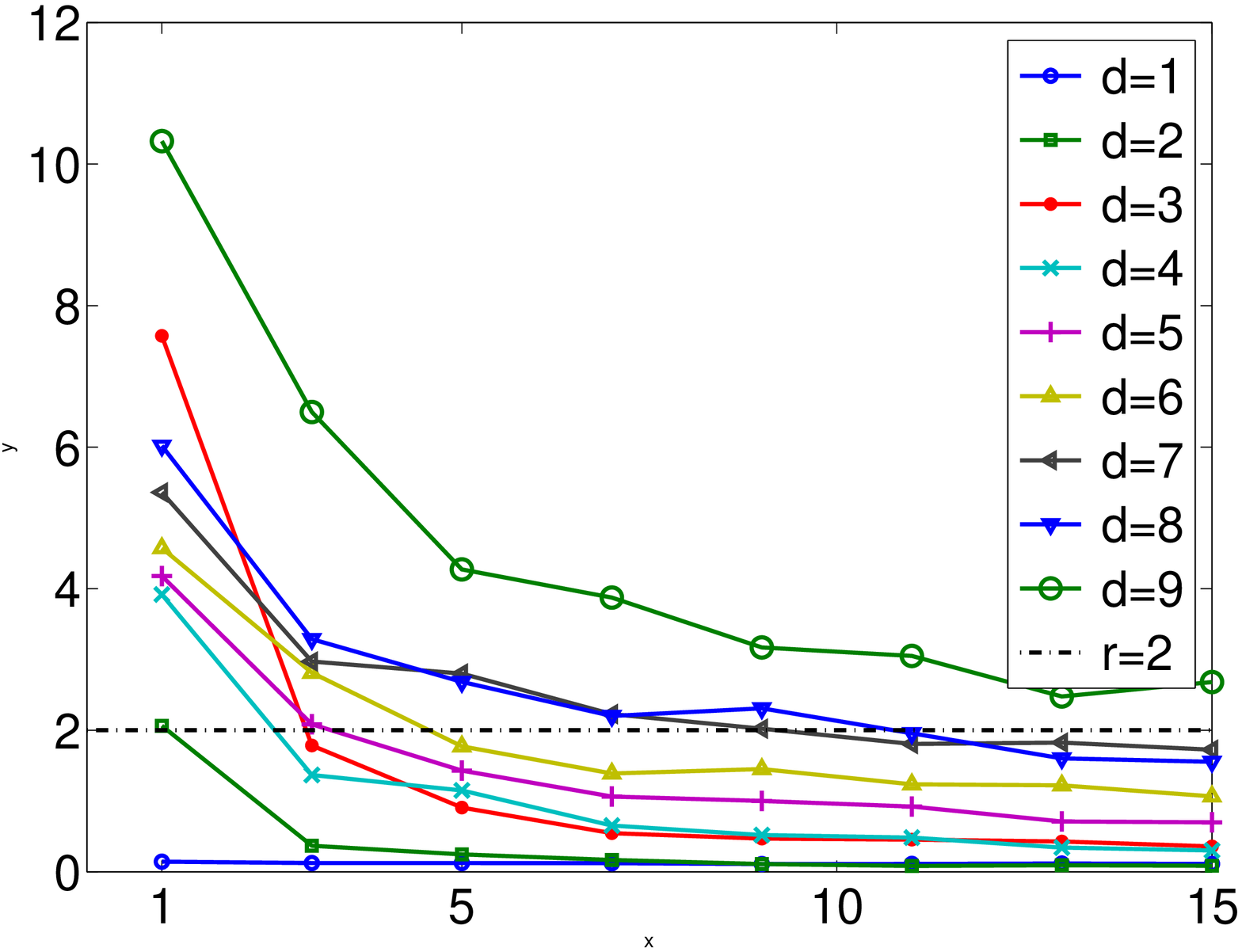}
   \caption{Mean dimension-free error as a function of the umber of Gaussian components $m$ for the \ac{BCF} algorithm. Each curve corresponds to a dimension $d$ and the horizontal line shows the fixed value of $r=2$.} 
   \label{fig:BCdaum}
\end{figure}

\begin{figure}[t]
	\centering
	\psfrag{x}[tc][][0.8]{Number of particles - $n$}
	\psfrag{y}[b][][0.8]{Mean dimension-free error - $r$}
	\ \\[0.06em]
	\hspace*{-3mm}
	\includegraphics[width=\columnwidth,keepaspectratio]{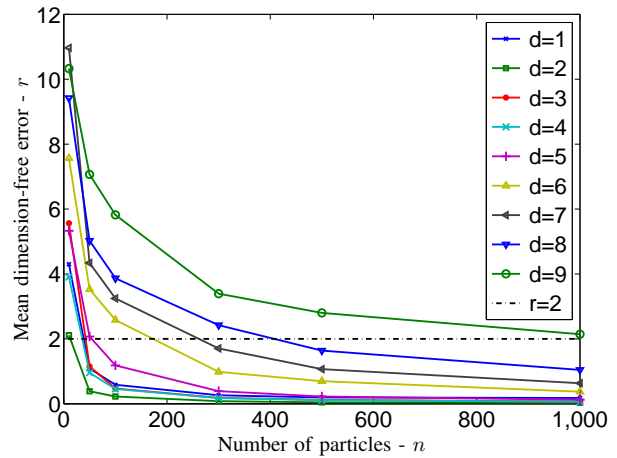}
   \caption{Mean dimension-free error as a function of the number of particles $n$ for the \ac{PF} algorithm. Each curve corresponds to a dimension $d$ and the horizontal line shows the fixed value of $r=2$.} 
   \label{fig:PFdaum}
\end{figure}

Figure \ref{fig:BCdaum} and Figure \ref{fig:PFdaum} show the Mean Dimension-Free Error plots for the \ac{BCF} and \ac{PF} respectively, one curve per dimension $d$. It can be seen that, for each curve, the error measure decreases with diminishing return as a function of increasing number of \ac{BCF} components and number of particles. This is in agreement with the \ac{MSE} analysis from Figure \ref{fig:BC_m} and Figure \ref{fig:PF_n} where accuracy improves with better parameters. Here, we set the Dimension-Free Error to be $r=2$ as shown by the black dotted horizontal lines. Where the curves cross the$r=2$ line specify the choice of $m$ and $n$ for for each dimension. For higher values of $d$, the intersecting points are moved further right, corresponding to larger component and ensemble sizes. This is expected, implying increased computation time as the dimensionality of the problem grows.

\begin{figure}[t]
	\centering
	\psfrag{x}[tc][][0.8]{Dimension ($d$)}
	\psfrag{y}[b][][0.8]{Computation time (sec)}
	\ \\[0.06em]
	\hspace*{-3mm}
	\includegraphics[width=\columnwidth,keepaspectratio]{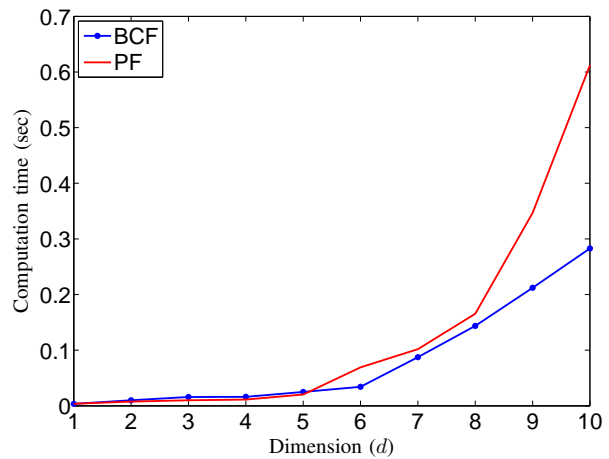}
   \caption{Complexity analysis in terms of computation time as a function of dimension ($d$) for the \ac{BCF} and \ac{PF} algorithms at $r=2$.} 
   \label{fig:complexity}
\end{figure}

Figure \ref{fig:complexity} shows the computational time for each filtering method as a function of varying state dimension following the above process above. From the figure, the \ac{PF} at very low dimensions shows a fast start, laying below the \ac{BCF} curve with only a few hundred particles. As the size of the problem increases, the computational time for \ac{PF} grows very rapidly, surpassing its counter part after $d=5$ from the numerical results. At dimension $d=10$ alone, with only a few thousand particles it can be seen that \ac{PF}'s computation time has doubled that of \ac{BCF}'s.

\begin{figure}[t]
  \centering
  
         \psfrag{x}[c][][0.8]{Number of particles - $n$}
	\psfrag{y}[b][][0.8]{Time (seconds)}
	\ \\[0.06em]
	\hspace*{-3mm}
	\includegraphics[width=\columnwidth,keepaspectratio]{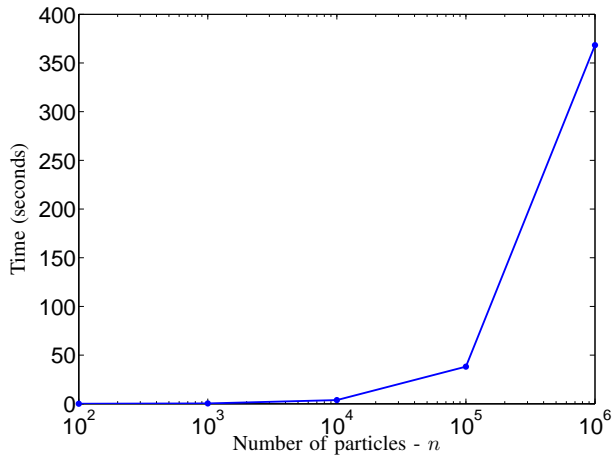}

  \caption{Operation time of \ac{PF} update step as a function of the size of the state vector.}
  \label{fig:time_n}
\end{figure}

Figure \ref{fig:time_n} shows the elapsed time for a \ac{PF} update as a function of the number of particles upto $n=10000$ for the IEEE 14-bus example . Here, the state is changed according to the dynamic behavior described above. Consider the case in which we are interested in tracking a household line voltage at \ac{SCADA} rate of 30 samples/second. This requires a sample roughly every $33 ms$. Theoretically, as the number of particles approaches infinity, the \ac{PF} can represent probability densities of any form no matter how complex they get. Unfortunately, it is evident that working with the \ac{PF} with $n>10000$ encounters major practical challenges.

\section{Conclusion} \label{chap4}

Power generation and distribution is quickly moving from a traditional top-down structure to smart grids, where power injection into and from the grid can occur at any point in the network. In order to ensure flawless operation through this paradigm shift, the need for \ac{SE} across the network with higher accuracies than ever before is inarguable. 

Classical approaches cast \ac{SE} as a \ac{HMM} whereby the state of the system is first predicted based on a priori knowledge (Prediction Step) and then the prediction is corrected using measurements (Update Step). The full system can be described by a forecasting equation capturing the time-varying dynamics of the system (Dynamic Model) and an observation function relating the state predictions to network measurements (Observation Model). There are four challenges that stand in the way of any algorithmic solution: 1) Scalability; 2) Integrated Dynamics; 3) Non-linear System Characteristics; 4) Uncharted System Noise. These factors are highly correlated and essential to the performance of \ac{SE} algorithms.  


The state-of-the-art filtering methods include \ac{LSQ} (widely adopted) and different variations of the Kalman filter (\ac{UKF} being the most up-to-date). The key assumptions that sit at the heart of deficiencies in current methodologies are the linear treatment of measurement function  $h(\cdot)$ and the Gaussian simplification of the noise model $\bf{n}_t$. The fundamental questions raised in this work are: 1) Can the non-linearities in the system be more realistically represented? 2) Can more general noise models be utilized to capture the network uncertainty?

This work puts forth two filtering methods, namely the \ac{BCF} and the \ac{PF}, which have successfully been used in other domains, and applies them for the first time to the \ac{SE} problem in power grids. We remove linearity constraints on the Observation Model and represent the complex power equations in their original form. Furthermore, we introduce a mixture uncertainty model that enables modeling of non-Gaussian noise in the system. Based on simulations of the IEEE 14-bus test-bench, numerical analysis in this work reveals that under system non-linearities and non-Gaussian noise, the \ac{BCF} and \ac{PF} achieve significant improvement in performance compared to \ac{LSQ} and \ac{UKF}. 

Simulations are run for different parameter settings of the proposed algorithms. In particular, the \ac{MSE} is used as a measure of accuracy for the \ac{BCF} with $m=\{ 1, 2, \cdots, 9\}$ components and for the \ac{PF} with upto $n=10,000$ particles. Under i.i.d additive Gaussian noise, even the baseline \ac{BCF} ($m=1$) and a small ensemble of particles ($n=1000$) reduce the \ac{MSE} over the \ac{UKF}, which is a second-order Kalman Filter. Increasing the number of components in the \ac{BCF} or particles in the \ac{PF} only improves the results. With $m=9$ components and $n=10,000$ particles, an \ac{MSE} per unit per line of $1.93\times10^{-4}$ and $1.2\times10^{-4}$ are achieved by each algorithm, in contrast to the \ac{UKF}'s \ac{MSE} per unit per line of $9.66\times10^{-4}$. For a network of size 14, this corresponds to net errors of $310 mV$ (the \ac{BCF}) and $180 mV$ (the \ac{PF}) versus $1.5 V$ (the \ac{UKF}) - an order of magnitude improvement by the \ac{BCF} and \ac{PF}. The non-Gaussianness in the system only makes these advantages more prominent. Under the most uncertainty (i.e. uniform noise), the \ac{MSE}'s increase to $1.4 \times 10^{-3}$ for the \ac{PF} with $n=10,000$, $1.6 \times 10^{-3}$ for the \ac{BCF} with $m=1$, and $4.5 \times 10^{-3}$ per unit per line for the \ac{UKF}.

The treatment of observation model as an inverse problem by the \ac{BCF} and the ensemble representation of the posterior density function by the \ac{PF}, make them an ideal choice for the problem of state estimation in power girds. The numerical analysis in this work show enhanced performance of at least an order of magnitude for these methods over the current technology - a difference that becomes crucial at scales and dynamic complexities required by future grids. Although the \ac{PF} itself can achieve better accuracies than the \ac{BCF}, this gain comes at a price. As the dimension of the problem at hand increases, numerical results in this work have shown that the computational complexity of the \ac{PF} increases in an almost exponential fashion, more than doubling in computation time at dimension $d=10$. Therefore, where computational cost is an important factor, the \ac{BCF} proves itself to be an ideal candidate - coming close to \ac{PF}'s accuracy while offering promising computing times to meet the needs of future Smart Grids.


\appendices

\section{Notation}

\begin{tabular}{ l l }
 $N$   & Number of bus lines   \\
 $ \theta _i$ & Phase angle at bus $i$ \\
 $ |V_i| $ & Voltage magnitude at bus $i$ \\
  $ I_{ij} $ & Current phasor on line between nodes $i$ and $j$ \\
 $P_i$ & Real Power at bus $i$ \\
 $Q_i$ & Reactive Power at bus $j$ \\
$ g_{ij}$ & Conductance between bus $i$ and $j$\\
$ b_{ij}$ & Susceptance between bus $i$ and $j$\\
$ g_{i0}$ & Shunt conductance at bus $i$ \\
$ b_{i0}$ & Shunt susceptance at bus $i$ \\

$\mathbf{x}_t$ & State Vector at time $t$ \\
$\mathbf{y}_t$ & Observation Vector at time $t$ \\
$g(\cdot)$ & Dynamic Model \\
$ \mathbf{q}_t$ & Dynamic Model Parameter at time $t$ \\
$h(\cdot)$ & Observation Model \\ 
$\mathbf{n}_t$ & Observation Noise at time $t$ \\

$\mathfrak{Re}\{ \cdot \}$ & Real part of a Complex Number \\
$\mathfrak{Im}\{ \cdot \}$ & Imaginary part of a Complex Number \\

$f(\cdot)$ & Probability Density Function \\
$\mathbb{ E }\{ \cdot \}$ & Expectation Operator \\
$  \mathcal{N}(\mathbf{\mu}, \mathbf{\Sigma}) $ & Gaussian Distribution \\ 

$\delta (\cdot)$ & Dirac delta function \\

$p_e$ & noise mixing coefficient \\

$m$ & Number of components in the \ac{BCF} \\
$n$ & Number of priceless in the $\ac{PF}$ \\
$q$ & Number of quadrature points \\
$d$ & The dimension of the state vector ($d \leq N$)\\

\end{tabular}

\section{Holt's Smoothing Method} \label{Appx:Holtz}

Holt's linear smoothing function is given by \cite{MakWhe:78}:
\begin{align}
\mathbf{x}_{t+1} = \mathbf{F}_t \mathbf{x}_{t} + \mathbf{g}_t + \mathbf{q}_t
\end{align}
where $\mathbf{q}_t$ is a white Gaussian sequence with zero mean and some covariance matrix, $\mathbf{F}_t $ is a nonzero diagonal matrix with dimension $n \times n$ and $ \mathbf{g}_t $ is a nonzero vector with dimension $n \times 1$. They are computed by following Holt's updates at each time $t$:
\begin{align}
\mathbf{F}_t &= \alpha_t (1+ \beta_t) . \mathbf{I} \\
\mathbf{g}_t &= (1+ \beta_t)(1 - \alpha_t) \mathbf{x}_{t}^{-}  - \beta_t \alpha_{t-1} + (1- \beta_t) \mathbf{b}_{k-1}
\end{align}
where  $\mathbf{I}$ is the identity matrix, and $\alpha_t$ and $\beta_t$ are constants between $0$ and $1$. Also, $\mathbf{x}_{t}^{-}$ here denotes the prediction for the state vector at time $t$, before updating with the measurement information. Further, the vectors $\mathbf{a}$ and $\mathbf{b}$ are computed through the recursions:
\begin{align}
\mathbf{a}_t &= \alpha_t  \mathbf{x}_t + (1-\alpha_t) \mathbf{x}_t^{-} \\
\mathbf{b}_t &= \beta_t ( \mathbf{a}_t  + \mathbf{a}_{t-1}  )+ (1- \beta_t) \mathbf{b}_{t-1}
\end{align}




\bibliographystyle{IEEEtran}
\bibliography{IEEEabrv,main}

\end{document}